\documentclass{aa}
\usepackage{graphicx,times}
\newlength{\fsize}
\def\Real{{\rm I\mathchoice{\kern-0.70mm}{\kern-0.70mm}{\kern-0.65mm}%
  {\kern-0.50mm}R}}  
  \def\bx#1{\leavevmode\thinspace\hbox{\vrule\vtop{\vbox{\hrule\kern1pt
  \hbox{\vphantom{\tt/}\thinspace{\bf#1}\thinspace}}
  \kern1pt\hrule}\vrule}\thinspace}

\def\be{\begin{equation}} \def\ee{\end{equation}}

\begin{document}

\title{Stable magnetic fields in stellar interiors}

\author{J. Braithwaite \inst{1,3}\and \AA. Nordlund \inst{2,4}}
\institute{$~^1$  Max-Planck-Institut f\"ur Astrophysik, Karl-Schwarzschild-Stra{\ss}e 1,
Postfach 1317,  D--85741 Garching, Germany\\
$~~^2$  Department of Astrophysics, Niels Bohr Institute, Juliane Maries Vej 30, 2100 K{\o}benhavn {\O}, Denmark\\
\tt$^3$jon@mpa-garching.mpg.de, $^4$aake@astro.ku.dk}
\authorrunning{Braithwaite and Nordlund}
\titlerunning{Stable magnetic fields in stellar interiors}

\abstract{
We investigate the 50-year old hypothesis that the magnetic fields of
the Ap stars are stable
equilibria that have survived in these stars since their formation. With numerical
simulations we find that stable magnetic field configurations indeed appear to exist under
the conditions in the radiative interior of a star. Confirming a
hypothesis by Prendergast (1956), the
configurations have roughly equal poloidal and toroidal field strengths. We find that tori of 
such twisted fields can form as remnants of the decay of an unstable random
initial field. In agreement with observations, the appearance at the surface is an
approximate dipole with smaller contributions from higher multipoles, and the surface field
strength can increase with the age of the star. The results of this
paper were summarised by Braithwaite \& Spruit (2004).
}

\maketitle

\section{Introduction}
\label{sec:intro}

The peculiar A and B stars (Ap-Bp) are main-sequence stars with a strong surface 
magnetic field. The nature of these fields has been the subject of a debate that has
accompanied the development of astrophysical magnetohydrodynamics since the 1950's.
The two leading possibilities were the `fossil field' theory
(\cite{Cowling:1945}) and the core dynamo theory.
The fossil field theory appears to be supported by some observations, such as the very high
strength of the field in some stars, the apparent stationary state of the field and the wide
range of field strengths observed. 

The main problem of the theory has always been the difficulty of finding realistic
equilibrium configurations for star-like objects with the analytic methods available, and of
demonstrating the stability of such configurations. With increasing computing power,
important aspects of this problem are now accessible by purely numerical means. While
numerical results for the present problem cannot match the precision of analytic methods,
they are excellent at providing clues about the kinds of magnetic configurations that
might exist, or to estimate the likelihood that the hypothesised stable fields can exist at all. 

In this paper we present numerical results that make the fossil field theory very plausible,
by showing that field configurations of a well-defined type appear to develop naturally
by the decay of stronger unstable fields. We begin with a brief summary of the relevant
observational properties of the magnetic stars.

\subsection{History and properties of the Ap-Bp stars}
\label{sec:hist}

Maury (1897) noted that the spectrum of $\alpha^2$CVn (one of the 
brightest of this class, at magnitude 2.9) was peculiar, showing
unusual weakness of the K line and strength of the Si II doublet at
4128{\AA}. Variability of some of the lines was subsequently
discovered and Belopolsky (1913) measured the changes in intensity and
radial velocity of one of the lines (Eu at 4129{\AA}), finding a
period of 5.5 days. The photometric light curve was measured
(\cite{GutandPra:1914}) and similar behaviour was later found in other
Ap stars (for instance \cite{Morgan:1933} and \cite{Deutsch:1947}).

Only upon the discovery of variable magnetic fields
(\cite{Babcock:1947}) did any explanation of this interesting spectral
behaviour become possible. It was found that Ap stars have an
unusually strong magnetic field, with surface strengths ranging
from a few hundred to a few tens of thousand gauss. The variability of
the field can be most easily explained by imagining a static field not
symmetrical about the rotation axis; the spectral peculiarity is then
taken to be a consequence of the effect the magnetic field has on the
transport of chemical species.

Various techniques have been
developed to observe the magnetic field on the surface of
stars. Measurement of the circular polarisation of the spectral lines
is used to give an average (weighted towards the centre of the disc) of the line-of-sight
component of
the surface field, called the {\it longitudinal field} in the
literature. In some stars with slow rotation (and hence small Doppler
broadening) spectral lines are split into separate Zeeman components,
in which case an average over the disc of the modulus of the field can be
obtained: this is called the {\it field modulus}. If one were to do this on the Sun, one would find that the
longitudinal field were extremely small in comparison to the
modulus. This is because the field has a small scale structure, and the
positive and negative regions of the line-of-sight component cancel
each other out. If one makes this observation of a magnetic Ap star,
this is not the case -- implying
a large scale structure. It is our task to find an explanation for
the strong, large-scale fields of Ap stars.

In addition to the longitudinal field and the field modulus, two more
quantities can readily be measured: the {\it quadratic field} and the
{\it crossover field}.
The former quantity is approximately proportional to $(\langle B^2
\rangle + \langle B_z^2 \rangle)^{1/2}$, where $B_z$ is the
line-of-sight component, and the latter is given by
$v\sin i \langle xB_z \rangle$, where $x$ is the normalised distance
from the stellar rotation axis in the plane of the sky.
(\cite{Mathys:1995a}, \cite{Mathys:1995b}).
This set of `observables' can be used to model
the field on the stellar surface -- one constructs a model
whose free parameters are made to converge on a solution by finding
the point of minimum disagreement with observations. 

Various models for the field configuration have been tried. The simplest
assume an axisymmetric field, inclined with respect to the rotation axis 
(e.g. Landstreet \& Mathys 2000). More elaborate models are those of 
Bagnulo et al. (2002), which assumes a field with dipole and quadrupole 
components at arbitrary orientations, and the
point-field-source model of Gerth et al. (1997). One thing
that all these models have in common is that they fail, in many stars,
to describe the observations accurately. This implies a more complex
field structure than can be written as the sum of low-order spherical
harmonics. 
On the other hand it is often found that parameter space contains several 
$\chi^2$ minima so it is not clear which one of these configurations, if any,
represents reality. Despite this, many of the results obtained do seem
to be reasonably model-independent. A more sophisticated approach 
which can yield better results is that of {\it
Zeeman-Doppler Imaging} (see \cite{PisandKoc:2002}), which has as of
yet only been applied to a very small number of stars, owing
to the high quality of the spectra required.

It has been suggested that Ap stars are all above a certain age --
Hubrig et al. (2000a) placed Ap stars on the H-R
diagram and found none in the first 30\% of their main sequence
lifetimes. It is possible that some dynamo process only begins at a
certain time, perhaps as the size of the radiative core changes (an A
star has a radiative envelope and a convective core); it is
also possible that the Ap progenitor contains a strong field in its
interior which only appears at the surface at some evolutionary
stage. However, this result does contradict some earlier results
(\cite{North:1993}; \cite{Wade:1997})
which claim that Ap stars are distributed uniformly across the width
of the main sequence band.

The rotation period of Ap stars tends to be longer than in normal A
stars (\cite{BonandWol:1980}). Whether the young Ap progenitors 
betray their destiny though a similarly long period is unclear; authors 
on the subject have yet to reach a definite conclusion (see, for instance, 
\cite{Hubrigetal:2000b}) and await observations more numerous than 
have so far been undertaken.

Landstreet \& Mathys (2000) find that the magnetic axis of
slowly-rotating ($P>25$ days) Ap stars 
is overwhelmingly more likely to be close to the rotation axis than one 
would expect from a random distribution -- of their sample of $16$ stars,
$14$ have the two axes within $30^\circ$ of each other, the
other two between $30^\circ$ and $45^\circ$. This result,
which was obtained using the best-fit method with an
axisymmetric field model, is reassuringly confirmed by Bagnulo et al. (2002)
who use a field consisting of dipole and quadrupole at arbitrary
orientation. The rapidly-rotating ($P<25$ days) stars, however,
show no such alignment -- the statistics are consistent with a random
orientation of the magnetic axis in relation to the rotation axis.

\section{Nature of the magnetic field in Ap stars} 

The question of how the structure of a star can accommodate a magnetic field, 
and if it can survive on a time-scale as long as the the main sequence
lifetime has accompanied the early development of astrophysical MHD
(e.g. \cite{Cowling:1958}, \cite{Chandrasekhar:1961},
\cite{Roberts:1967}; the topic is reviewed in \cite{Borraetal:1982}).
The question has two parts: equilibrium and stability.

\subsection{Equilibrium} 

Finding equilibrium configurations of stars with
magnetic fields turns out to be a mostly technical problem. Though early
efforts, concentrating on `analytical solutions', had limited success, this does
not imply a conceptual problem affecting the existence of magnetic equilibria.
Construction of equilibria by numerical methods has become an accepted
approach (e.g. \cite{Bonazzolaetal:1993}).

For a dynamical equilibrium, there has to be a balance between the pressure
gradient, gravity and the Lorentz force. The Lorentz force is generally
not a conservative force, hence cannot be balanced by the pressure
gradient alone. Gravity, or more accurately buoyancy forces, must be
involved in maintaining equilibrium. For a given magnetic field configuration,
it is in general possible to find a (slightly distorted) stellar model that will 
balance the magnetic forces throughout the star. To see this (without actual
proof), note that the three components of a magnetic field can in general be
described in terms of two scalar fields (since ${\rm div} {\bf B}=0$ takes
care of one degree of freedom). Hence the magnetic force can also be
described in terms of two degrees of freedom only. Ignoring thermal
diffusion, the thermodynamic state of the gas has two degrees of freedom
(pressure and entropy, for example). Where the magnetic field is sufficiently
weak (in the sense $B^2/(8\pi P)\ll 1$), equilibrium therefore can be
obtained, for a given magnetic configuration, by suitable small adjustments of
the pressure and entropy distributions. An exception occurs in convectively
unstable layers, which do not support significant differences in entropy.

Where the field strength is not small in this sense, for example in the atmosphere
of the star, not all field configurations are possible, and the magnetic field
must instead be close to a  force-free configuration. Conceptually, we can
thus divide a radiatively stratified star into an interior where any field
configuration is allowed (if adiabatic equilibrium is the only concern, and up
to some maximum strength), and an atmosphere containing a nearly force-free 
field. The two join somewhere around the surface where $B^2/(8\pi P)=1$.

Deviations from magnetic equilibrium travel through the star at the Alfv\'en
speed. Even if the magnetic field in the interior is weak ($B^2/(8\pi P)\ll
1$), the corresponding adjustment time can still be very short compared
to the age of the star, since this is so many orders of magnitude longer
than the dynamical time-scale of a star (of the order $10^{11}$ times longer,
for a main sequence A-star). For a field of 1000 G in an Ap star, for
example, the Alfv\'en crossing time is of the order 10 years, a fraction
$10^{-8}$ of the star's main sequence life.

In our calculations, the laborious process of producing magnetic equilibria
by explicit construction from the equilibrium equations is replaced by the `brute
force' method of following the evolution of the configuration in a time dependent
manner. Though less elegant, it is simpler to implement and has the additional 
advantage of addressing at the same time the stability of the field.

\subsection{Stability} The (dynamic) stability of an equilibrium is equally
important, since instability will result in changes on the same
(Alfv\'en) time-scale. Gravity (buoyancy) is a strongly stabilising force on the field in a
radiative stellar interior, preventing displacements in the radial direction.
But in the two horizontal directions (along an equipotential surface), there is
essentially no stabilising force. Is stabilisation in one direction sufficient for
overall stability of magnetic equilibria in stars? What do such equilibria look
like if they exist? This question has been the subject of a significant amount 
of analytic work done throughout the last fifty years.

Tayler (1973) looked at toroidal fields in stars, that is, fields that have only an
azimuthal component $B_\phi$ in some spherical coordinate frame $(r,\theta,\phi)$
with the origin at the centre of the star. 
With the energy method, he derived necessary and sufficient stability conditions 
for adiabatic conditions (no viscosity, thermal diffusion or magnetic diffusion). 
The main conclusion was that such purely toroidal fields are always
unstable at some place in the star, in particular to perturbations of the $m=1$ form, 
and that stability at any particular place does not depend on field strength but 
only on the form of the field. 
An important corollary of the results in this paper (esp. the Appendix) was
the proof that instability is {\it local} in meridional planes. If the necessary 
and sufficient condition for instability is satisfied at any point $(r,\theta)$,
there is an unstable eigenfunction that will fit inside an infinitesimal environment
of this point. The instability is always global in the azimuthal direction, however. 
The instability takes place in the form of a low-azimuthal order displacement in 
a ring around the star. Connected with this is the fact that the growth time of
the instability is of the order of the time it takes an Alfv\'en wave to travel 
around the star on a field line.

\begin{figure}
\includegraphics[width=1.0\hsize,angle=0]{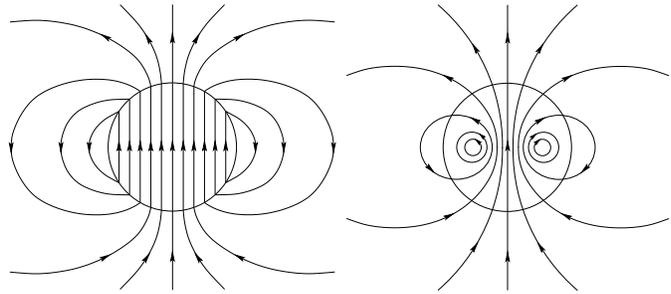}
\caption{Poloidal field configurations. Left: all field lines close outside the star,
this field is unstable by an argument due to Flowers \& Ruderman. For the case
where some field lines are closed inside the star, instability was proven by 
Wright and Markey \& Tayler.}
\label{fig:initfield}
\end{figure}

The opposite case is a field in which all field lines are in meridional planes 
($B_\phi=0$, see Fig. \ref{fig:initfield}). In subsequent papers Markey \& 
Tayler (1973, 1974) and independently Wright (1973) studied the stability of 
axisymmetric poloidal fields in which (at least some) field lines are closed 
within the star (right-hand side of Fig. \ref{fig:initfield}). These
fields were again found to be unstable.

A case not covered by these analyses was that of a poloidal field in which 
{\it none} of the field lines close within the star. An example of such a field is
that of a uniform field inside, matched by a dipole field in the vacuum outside 
the star (left-hand side of Fig. \ref{fig:initfield}). This case has been considered earlier by Flowers \& 
Ruderman (1977) who found it to be unstable, by the following argument. 
Consider what would happen to such a dipolar field if
one were to cut the star in half (along a plane parallel to the
magnetic axis), rotate one half by $180^\circ$, and put the two halves
back together again. The magnetic energy inside the star would be
unchanged, but in the atmosphere, where the field can be approximated
by a potential field, i.e. with no current, the magnetic energy will
be lower than before. This process can be repeated {\it ad infinitum} --
the magnetic energy outside the star approaches zero and the
sign of the field in the interior changes between thinner and thinner
slices.

The reduction of the external field energy is responsible for
driving the instability. Since the initial external field energy is of the
same order as the field energy inside the star, the initial growth time
of the instability is of the order of the Alfv\'en travel time through the
star, as in the cases studied by Markey \& Tayler and Wright. 

Prendergast (1956) showed that an equilibrium can be constructed from
a linked poloidal-toroidal field, but stopped short of demonstrating
that this field could be stable. Since both purely toroidal fields and
purely poloidal field are unstable, a stable field configuration, if
one exists, must apparently be such a linked poloidal-toroidal
shape. Wright (1973) showed that a poloidal field could be stabilised by 
adding to it a toroidal field of comparable strength. However, the result 
was again somewhat short of a proof.

Kamchatnov (1982) constructed an equilibrium field, which he claimed
was stable. It has the following form:

\begin{eqnarray}
B_x & = & \frac{2(xz-y)}{(1+r^2)^3}\nonumber\\
B_y & = & \frac{2(yz-x)}{(1+r^2)^3}\\
B_z & = & \frac{1+2z^2-r^2}{(1+r^2)^3}\nonumber
\label{eq:kamfield}
\end{eqnarray}

where $x$, $y$ and $z$ are unitless Cartesian coordinates, and
$r^2=x^2+y^2+z^2$. To be in equilibrium, this field has to be accompanied by a
velocity field of similar form. The field is a twisted torus; if one
started with a hoop of field lines (i.e. a toroidal field), cut the
hoop at one point, and twisted one end $360^\circ$ and reconnected the
two ends again (so that each field line connects back to itself), one would
get something like the field described by the equations above.

These results were all valid only in the absence of dissipative effects. 
The damping due to such effects might be expected to result in a
somewhat increased stability. The only case in which dissipative
effects have been investigated in detail is that of a purely toroidal field.
Acheson's (1978) analysis exploits the local nature of the instability 
process in this case to include the effects of viscosity, magnetic and
thermal diffusion. Because of the low values of these coefficients
in a stellar interior, stabilisation is found to occur only at very low field 
strengths, well below those observed in Ap/Bp stars.

The effect of rotation was investigated by Pitts \& Tayler (1986) for the
adiabatic case. These authors arrived at the conclusion
that although some instabilities could be inhibited by sufficiently
rapid rotation, other instabilities were likely to remain, whose
growth could only be slowed by rotation -- the growth timescale would
still be very short compared to a star's lifetime. It seems that a
toroidal field could be stabilised by rotation above a certain speed
if the rotation and magnetic axes were parallel. However, there are
generally likely to be other instabilities which survive even rapid
rotation, albeit at a rate reduced by a factor $\sigma_0/\Omega$ where
$\sigma_0$ is the growth rate in the absence of rotation. They did not
however exclude the possibility that rotation at a large angle to the
magnetic axis of symmetry could stabilise a mainly poloidal field.

This was one of the last papers on the subject to use purely analytic
methods -- the problem had become so complicated that no more definite
conclusions could be made.
Numerical simulations were recently used to look at
the stability of toroidal fields (\cite{Braithwaite:2004}); it was 
demonstrated that such a field is subject to an
instability growing on an Alfv\'en-crossing time-scale, which could be
suppressed by rotation of an axis parallel to the magnetic axis. These
simulations were done in a localised basis -- a small section of the
radiative envelope on the magnetic axis was modelled. To look at the
stability of more general field configurations, it is necessary to
model an entire star.

In the calculations reported below the stability problem is not studied 
separately; any configuration that survives the dynamical evolution
of a given initial state will be a stable field, on the time-scales
that can be followed numerically. The evolution can typically be followed
for a few hundred Alfv\'en crossing times; surviving fields are therefore
of the dynamically stable type sought.

It is possible that the outcome depends on the initial conditions, which 
could of course explain why some A stars are magnetic and others are 
not. A second goal is thus to find clues as to the initial conditions set at the
time of formation of a magnetic A star.

\section{The numerical model}
\label{sec:model}

The star is modelled on a Cartesian grid. For a spherical object like
a star this  might sound unnatural. Alternatives like cylindrical or
spherical coordinates are more natural for analytical methods, but are
known to produce serious artefacts in numerical simulations because of
the coordinate singularities. Cartesian coordinates are the simplest
to implement and have a low computing cost per grid point. A
disadvantage is that the  computational box must be taken somewhat
larger than the star studied,  which increases computing effort again.

The boundary conditions used are periodic in all directions. Such
conditions are easy to implement and minimise boundary artefacts.

The equation of state is that of an ideal gas with a fixed ratio of
specific heats $\gamma=5/3$. The gravitational potential is determined
consistently with the non-magnetic state of the star, but thereafter
kept fixed at this value during the  evolution of the magnetic field
(the Cowling approximation).

It should be stressed that the star modelled here is
non-rotating. This is no problem for slowly-rotating Ap stars where
the Alfv\'{e}n time-scale is much shorter than the rotational
time-scale; in the faster-rotating Ap stars, since rotation tends to
suppress magnetic instabilities, the effect of the rotation on any
stable field configuration is likely to be one of orientation
only. This is confirmed by the observations mentioned in the
introduction (\cite{LanandMat:2000}) -- that the only difference
between the slow-rotators and the fast-rotators is the angle between
the rotation and magnetic axes.

\subsection{Treatment of the atmosphere}
\label{sec:apfa}

As the Flowers-Ruderman argument shows, instability of the field can
be driven by the magnetic energy in the volume outside the star. The 
calculations therefore must include a mechanism to allow magnetic energy 
release in the atmosphere. The atmosphere is magnetically dominated 
($\beta\ll1$) and has something close to
a potential field, as no large currents can exist there. 
In principle, the code will reproduce this automatically, since magnetic diffusion
(whether numerical or explicitly included) will allow reconnection of field lines
in the atmosphere in response to changes at the surface driven by the dynamics
of the magnetic field in the interior.

When numerically modelling this, however, problems arise because the
Alfv\'en speed becomes very large in an atmosphere that is modelled
sufficiently realistically to allow reconnection to take place rapidly
enough. This causes the time step to drop below acceptable values.

By including a large electrical
resistivity in the atmosphere, the field can be kept close to a 
potential field irrespective of the Alfv\'en speed. 
Thus in the induction equation 
a magnetic diffusivity, $\eta_a$, is included, whose value is 
zero in the stellar interior and constant in the atmosphere
(with a transition zone located between the same radii as the
temperature transition zone visible in Fig. \ref{fig:profile-temp}). 
The corresponding heating term in the energy equation is left
out since the diffusivity is artificial, and atmospheric heating can not 
be treated realistically anyway without also including the compensating
radiative loss terms.

\subsection{Time-scales and computational practicalities}
\label{sec:tscales}

Three different time-scales play a role in the numerics of the problem: 
the sound travel time $t_\mathrm{s}=R_\ast/c_\mathrm{s}$, the Alfv\'en 
crossing time $t_\mathrm{A}= R_\ast/v_\mathrm{A}$ and the Ohmic diffusion 
time $t_\mathrm{d}=R_\ast^2/\eta$. In a real star (assuming a field strength
$\sim 1000$ G), these differ by ratios of the order $t_\mathrm{A}/
t_\mathrm{s}\sim 10^{4}, t_\mathrm{d}/ t_\mathrm{s}\sim 10^{10}$. 
Such ratios are well outside the dynamic range accessible numerically.

In the main problem addressed in this study, namely the approach of a field 
configuration to an equilibrium state and the stability of this state, the 
governing time-scale is the Alfv\'en travel time. The hydrostatic adjustment 
of the star to a changing field configuration happens on the much shorter sound 
crossing time, hence the evolution of the field does not depend explicitly on the
sound crossing time, but only on the Alfv\'en speed. An overall change in the
field strength is thus almost equivalent to a change in time scale. We exploit 
this by using high field strengths, such that $t_\mathrm{A}/t_\mathrm{s} 
\sim 10$, a value close to the maximum for which the dynamics can plausibly be expected
to be nearly independent of the sound travel time.

In some calculations, the evolution of a dynamically stable field configuration 
in the presence of an explicit magnetic diffusion is studied (see
Sect.~\ref{sec:phase2}). In these cases, both the dynamic and the diffusive time-scales 
must be followed. The two can be separated only if the diffusive time-scale is 
sufficiently long compared with the Alfv\'en time. For these cases, the diffusivity 
is adjusted such that the diffusive time scale was longer by a factor of order 10; 
hence these calculations are also of the order 10 times as demanding as the
calculations that only follow the Alfv\'en time scale.

\subsection{Acceleration of the code by rescaling}
\label{sec:resca}
During the evolution of the field from an arbitrary or unstable configuration,
its amplitude decreases by large factors. Following the intrinsic development 
as the Alfv\'en time-scale increases becomes increasingly expensive, limiting
the degree of evolution that can be followed. To circumvent this problem, a 
routine was added to the code which increases the strength of the magnetic 
field (uniformly throughout the entire computational box) to keep the total 
magnetic energy constant. The code then keeps a record of how fast the 
magnetic field would have decayed in the absence of this routine. This 
information is then used to reconstruct the time axis and the field amplitude 
as a function of time. 
We call this numerical device {\it amplitude rescaling}. 
It can be shown to give exact results in the case when the Alfv\'en speed is 
the only relevant signal speed. In practice, this means we expect it to give 
a good approximation for the evolution of the field configuration when the 
Alfv\'en crossing time is much longer than the sound crossing time but much 
shorter than the diffusive time, i.e. in the limiting case $\eta/R_\ast \ll V_{\rm A}\ll 
c_{\rm s}$.  Tests were done to make sure that the evolution of the field is 
indeed largely unaffected by this procedure (Sect. \ref{sec:test}.)
The procedure is useful even in cases where the separation of Alfv\'en, sound and
diffusive time scales is not as clean. If we are interested mainly in the stable {\it final
equilibrium configuration}, it is sufficient to have a numerical procedure that finds
this equilibrium efficiently and the accuracy of the evolution to this equilibrium is 
of less concern. This applies in essence to most of the results reported here.

The model does not include thermal processes such as production of heat in the core and transport to the surface, and Ohmic heating from electrical currents; the latter would be nonsensical anyway as a result of the rescaling of the magnetic field. Because of this, and because the star is not rotating, we do not expect the results to give us useful information about flows in the stellar interior.

\section{The numerical code}
\label{sec:code}

We use a three-dimensional MHD code developed by Nordlund \&
Galsgaard (1995), written in Cartesian coordinates.
The code uses a staggered mesh, so that variables are defined at
different points in the gridbox. For example, $\rho$ is defined in the
centre of each box, but $u_x$ in the centre of the face perpendicular
to the x-axis, so that the value of $x$ is lower by
$\frac{1}{2}dx$. Interpolations and spatial derivatives are calculated
to fifth and sixth order respectively. The third order
predictor-corrector time-stepping procedure of Hyman
(1979) is used.

The high order of the discretisation is a bit more expensive per grid point
and time step, but the code can be run with fewer grid points than 
lower order schemes, for the same accuracy. Because of the steep 
dependence of computing cost on grid spacing (4th power for explicit 3D) 
this results in greater computing economy.

For stability, high-order diffusive terms are employed. Explicit use
is made of highly localised diffusivities, while retaining the
original form of the partial differential equations.

The code conserves $\nabla.\mathbf{B}$ only up to machine accuracy. 
For previous applications this was no problem as the code was run for 
shorter lengths of time. For this application, however, we are modelling 
a star over many Alfv\'{e}n timescales, and accumulation of machine
errors became a problem. 
An additional routine was required to remove the component of the
field with non-zero divergence. This was done by periodically (every
few hundred timesteps) expressing the field as the gradient of a
scalar and the curl of a vector, the former then being deleted.



\section{Initial conditions}
\label{sec:init}

We begin all of the simulations with the spherically symmetrical density
and temperature profiles of polytropic star where $P 
\propto \rho^{1+1/n}$. A value of $n=3$ was chosen, as it is fairly
typical of the non-convective stellar envelope of A stars --
half-way between the isothermal ($n=\infty$) and convective
($n=3/2$) cases. This polytrope is truncated at a distance $R_\ast$,
the radius of the simulated star, and the region outside this replaced 
by a hot atmosphere, with a temperature about half that at the centre 
of the star. The surface density of the polytrope is of the order 
0.002 of the central density; a smaller value of the surface 
density is numerically impractical because of the very high Alfv\'en speeds 
that would result in the atmosphere.

If we choose to specify the mass $M_\ast$, radius $R_\ast$ and mean
molar mass $\mu$ of the star as $2M_\odot$, $1.8R_\odot$ and $0.6\,
\mathrm{g\,mol^{-1}}$ (typical A-star values), then the central
temperature is $9\times10^6$ K with a polytrope of this index. The
computational box is a cube of side $4.5R_\ast$.

\begin{figure}
\includegraphics[width=1.0\hsize,angle=0]{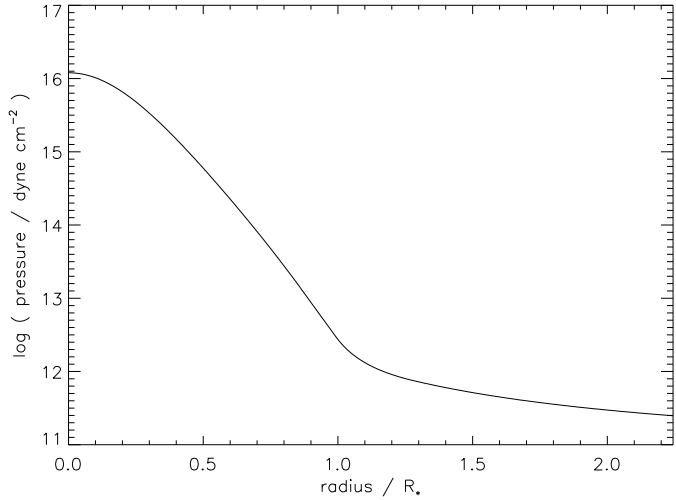}
\caption{Pressure vs. radius of the model star at $t=0$.}
\label{fig:profile-lnp}
\end{figure}

\begin{figure}
\includegraphics[width=1.0\hsize,angle=0]{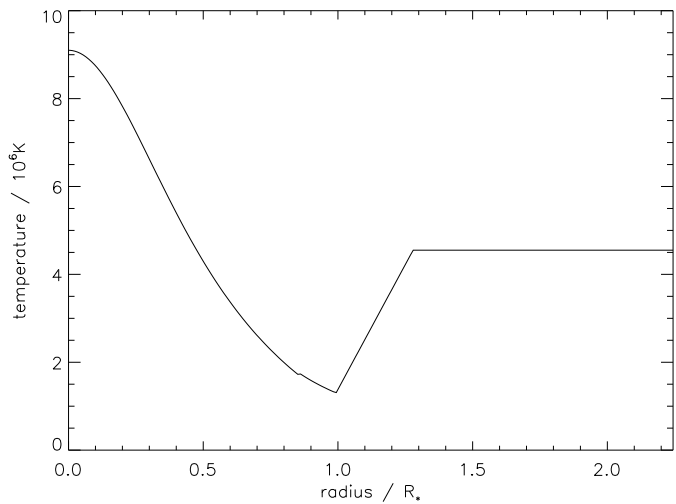}
\caption{Temperature  vs. radius of the model star at $t=0$.}
\label{fig:profile-temp}
\end{figure}

Figs. \ref{fig:profile-lnp} and \ref{fig:profile-temp} show the
pressure and temperature profiles.

Nothing is known about the magnetic configuration produced during star
formation. As a way of expressing this ignorance, have we started the series 
of calculations with a set of cases with random initial fields. 
A magnetic vector potential was set up as a random field containing spatial
frequencies up to a certain value (so that the length scales were at
least a few grid spacings). The result was then multiplied by 
\begin{equation}
\exp(-r^2/r_\mathrm{m}^2),
\label{eq:rm}
\end{equation}
so that the
field strength in the atmosphere was negligible. The magnetic field was
then calculated from this vector potential. The strength of the field was normalised to
be strong enough to allow things to happen on computationally
convenient timescales whilst holding to the condition that the
magnetic energy be much less than the thermal. The value of
$\beta$ (thermal over magnetic energy density) in the stellar interior was therefore set to
around $100$ at the beginning of the
simulation. The total magnetic energy is equal to
$1.2\times10^{46}$ erg -- this corresponds to a mean field of around
$5\times 10^6$ gauss, a factor of between $200$ and $20,000$ greater than the
fields observed on the {\it surface} of Ap stars. The Alfv\'{e}n
timescale, $0.6$ days, will therefore also be shorter by this factor
than one might expect in reality.

\begin{figure}
\includegraphics[width=1.0\hsize,angle=0]{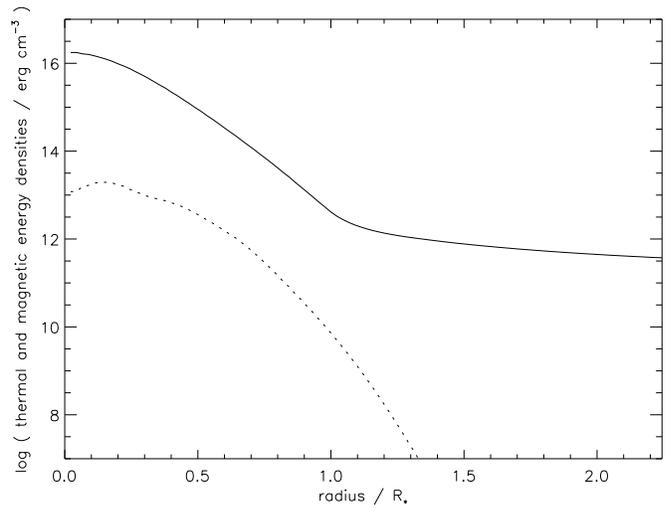}
\caption{Thermal (solid line) and magnetic (dotted line) energy densities 
at $t=0$, averaged over horizontal surfaces, as a function of radius. The 
variations of magnetic energy density with radius reflect the particular 
realisation of the random initial field.}
\label{fig:profile-enden}
\end{figure}

\begin{figure}
\includegraphics[width=1.0\hsize,angle=0]{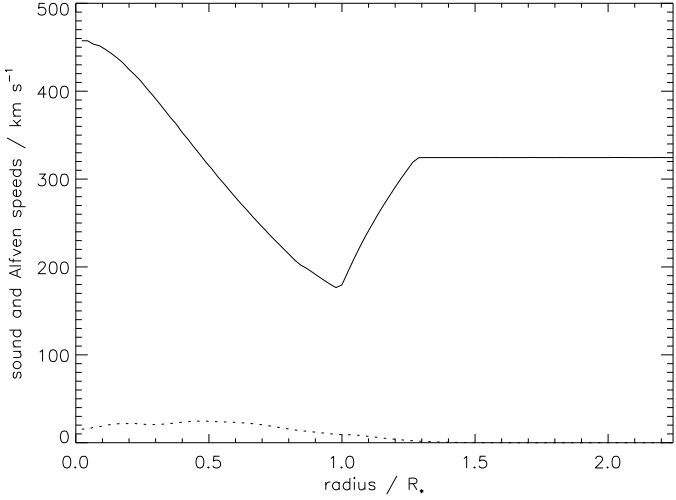}
\caption{Sound (solid line) and Alfv\'{e}n (dotted line) speeds at
$t=0$.}
\label{fig:profile-speeds}
\end{figure}

Fig. \ref{fig:profile-enden} shows the thermal
and magnetic energy densities as a function of radius for one 
particular realisation of the random initial conditions. The
difference between these two lines gives a measure of $\beta$, which
is typically $100$ in the interior, tending to infinity in the
atmosphere. Perhaps more relevant though than the ratio of the energy
densities is the ratio of sound and Alfv\'{e}n speeds; these two
speeds are shown in Fig. \ref{fig:profile-speeds}.

\section{Visualising the results}
\label{sec:vis}

One of the greatest difficulties in any numerical study of magnetic
fields is the visualisation of the results. A number of techniques
have been used in this study. One of the most useful pieces of
software was found to be \cite{Explorer}, which allows user-defined
procedures to easily be added to a pre-existing set of 3-D rendering
modules.  For most of the renderings shown here we use a combination
of user-defined and pre-existing modules that combines projections of
three-dimensional field lines and surfaces. A field line projection routine has been
used which picks starting points, either at random throughout the whole
computational box (biased in favour of regions of high field strength)
or around a certain point, and then traces the field from these points
until the field strength drops below a certain value.

Plotting field lines alone can produce a rather chaotic picture, as it
is not obvious at what depth each line lies. To complement the lines,
it is useful to plot an opaque surface of constant radius to provide a
background. This surface can then be shaded according to the sign of
the normal field component. A plot of this form can be seen in Fig.
\ref{fig:foursnaps} (the four large frames in this figure).

It is also helpful to be able to see the axis of symmetry of the
magnetic field. We define the magnetic axis $\mathbf{M}$ in the
following way:

\begin{equation}
\mathbf{M}  = \oint\limits_{r=R_\ast} (\mathbf{B.\hat{r}})\mathbf{r}
\,dS,
\label{eq:m1}
\end{equation}

The arrow representing this axis has
been added to the snapshots in Fig. \ref{fig:foursnaps}.

The stable magnetic field configurations found here generally have the 
form of tori. To visualise their shape, something is needed to show them 
as surfaces or nested surfaces. Field lines are needed too, but bundles of 
field lines alone are much too confusing, while iso-surfaces of field strength 
are too insensitive to the topological properties of the configuration.

A useful visual aid which helps to highlight the position of the
torus field is created as follows. A scalar field $C$ is calculated,
which is equal to the radius of curvature:
\begin{equation}
C = \frac{B^3}{|\mathbf{B\times A}|} \qquad \mbox{where} \qquad
\mathbf{A} = (\mathbf{B.\nabla})\mathbf{B} .
\label{eq:radofcurv}
\end{equation}
This alone is not the ideal
field to highlight the torus, as it fails to distinguish the core of
a torus from the field lines which go through the middle of the star
and emerge at either end. It is therefore necessary to look at the
direction of this radius of curvature -- we are interested principally
in places where it is parallel to the radius vector
$\mathbf{r}$. Hence the dot product of the unit position vector and
the radius-of-curvature vector $\mathbf{C}$ is calculated, giving a
scalar field $F$:
\begin{equation}
F = \mathbf{\hat{r}}.\mathbf{C} \qquad \mbox{where} \qquad
\mathbf{C} = \frac{B^2\mathbf{A}-(\mathbf{A}.\mathbf{B})\mathbf{B}}
                 {|B^2\mathbf{A}-(\mathbf{A}.\mathbf{B})\mathbf{B}|} C.
\label{eq:vecofcurv}
\end{equation}
We now wish to highlight regions where this scalar field $F$ is high
along a thin filament parallel to the magnetic field. We do by looking
at the second spatial derivatives perpendicular to the magnetic field:
\begin{equation}
G = -r^2\mathbf{r.\hat{C}}(\mathbf{\nabla\times B})^4
((\mathbf{m.\nabla})^2 F+(\mathbf{n.\nabla})^2 F)
\label{eq:finaltorusplot}
\end{equation}
where $\mathbf{m}$ and $\mathbf{n}$ are unit vectors perpendicular to
both $\mathbf{B}$ and each other. Adding the current density and radius
factors
both help to make the path of the torus stand out better. It is this
scalar field $G$ which is plotted in the smaller frames of Fig. \ref{fig:foursnaps}.

In addition to these plots, it is possible to project the surface of
the star into two dimensions. Enjoying the luxury of being subject to neither navigational
nor political considerations, we picked the simplest projection
imaginable, that is, longitude becomes the $x$ coordinate and latitude
becomes the $y$ coordinate. 
This is useful in particular when the field has a dominant component by
which an axis can be defined. In many of the configurations evolving found
here, a dipole component dominates at the surface, and its axis (the
axis $\mathbf{M}$ defined in Eq. (\ref{eq:m1})) is then taken
for the $(x,y)$-projection.

The shape and evolution of the magnetic field can be quantified in the following
ways. Firstly, as the length of the intersection between the $B_r=0$
and $r=R_\ast$ surfaces -- plotted in the following section is $W$, the
length of this intersection divided by $2\pi R_\ast$. A value of $1$ is
expected for a dipolar field; a value of five or more implies a field
with structure on the scale of a few grid points.

Secondly, the surface value of $B_r$ can be 
decomposed into spherical harmonics. This gives an indication
of how well ordered the field is -- if we plot the energy of the
dipole, quadrupole, octupole and higher orders as a fraction of the
total field energy, it is easy to see how ordered or chaotic the field
is. The coefficients can also be compared directly to the results of
observational studies which have assumed a dipole or dipole and
quadrupolar field. The axis $\mathbf{M}$ defined above is parallel to the dipole
moment.

Thirdly, we can break up the magnetic field into its three components
in spherical coordinates (again using the axis $\mathbf{M}$), and then calculate the total energy in the
toroidal component and in the poloidal component.

Finally, we calculate a radius $a_\mathrm{m}$ to quantify the 
volume occupied by the magnetic field:
\begin{equation}
a_\mathrm{m}^2=\frac{\int B^2 r^2 dV}{\int B^2 dV}.
\label{eq:def-am}
\end{equation}
This is especially useful for calculations of the longer-term diffusive
evolution of the field, giving a measure how far outwards the field has
spread from its initial form.
The initial value of $a_\mathrm{m}$ is roughly equal to the length scale 
$r_\mathrm{m}$ of the initial field configuration (Sect. \ref{sec:init}). It
will also depend to some small extent on the exact form of the initial
random field, in general $a_\mathrm{m}(t=0)\approx 0.9 r_\mathrm{m}$.

\section{Results}
\label{sec:results}

\begin{figure*}
\includegraphics[width=0.43\hsize]{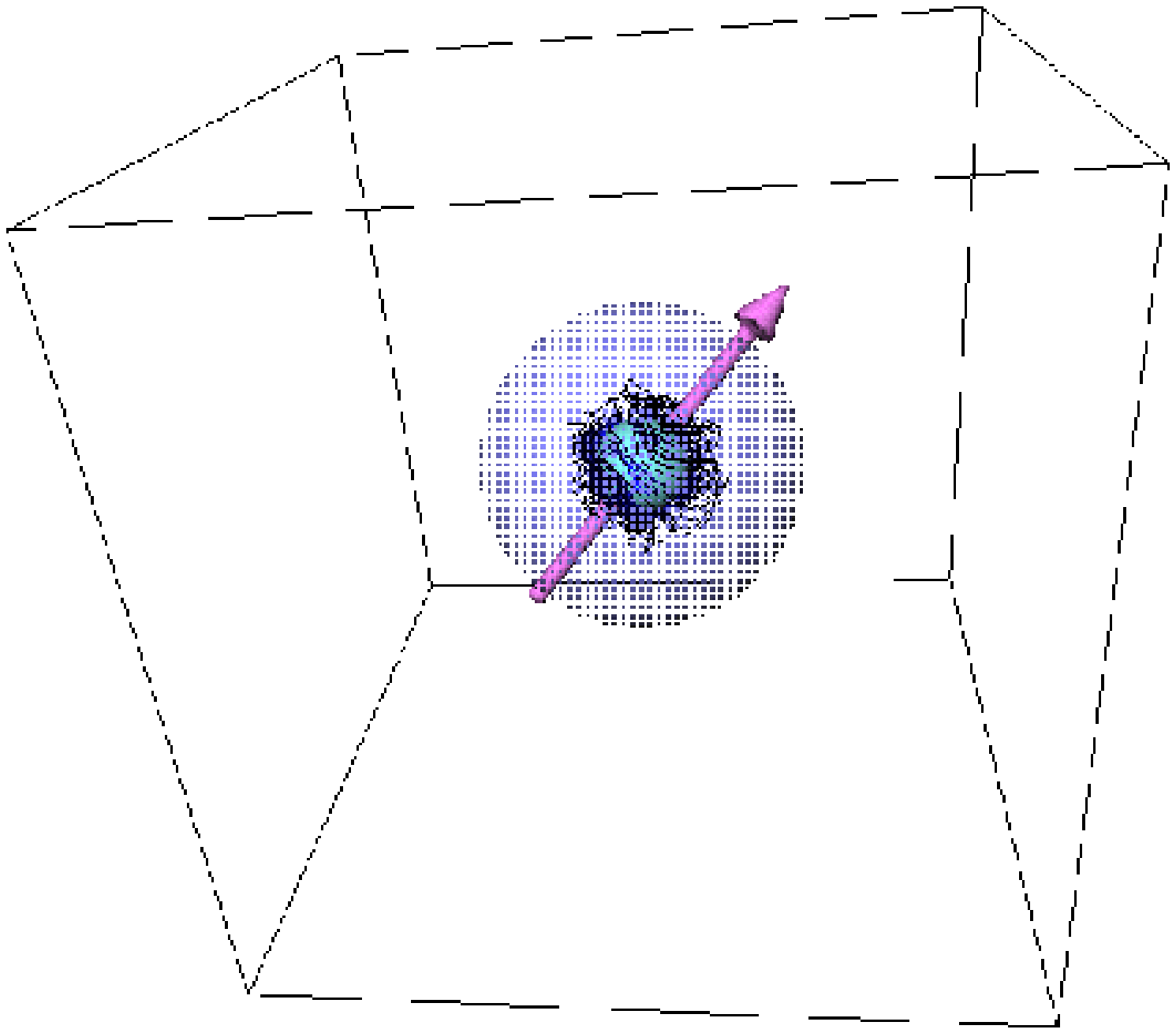}
\includegraphics[width=0.43\hsize]{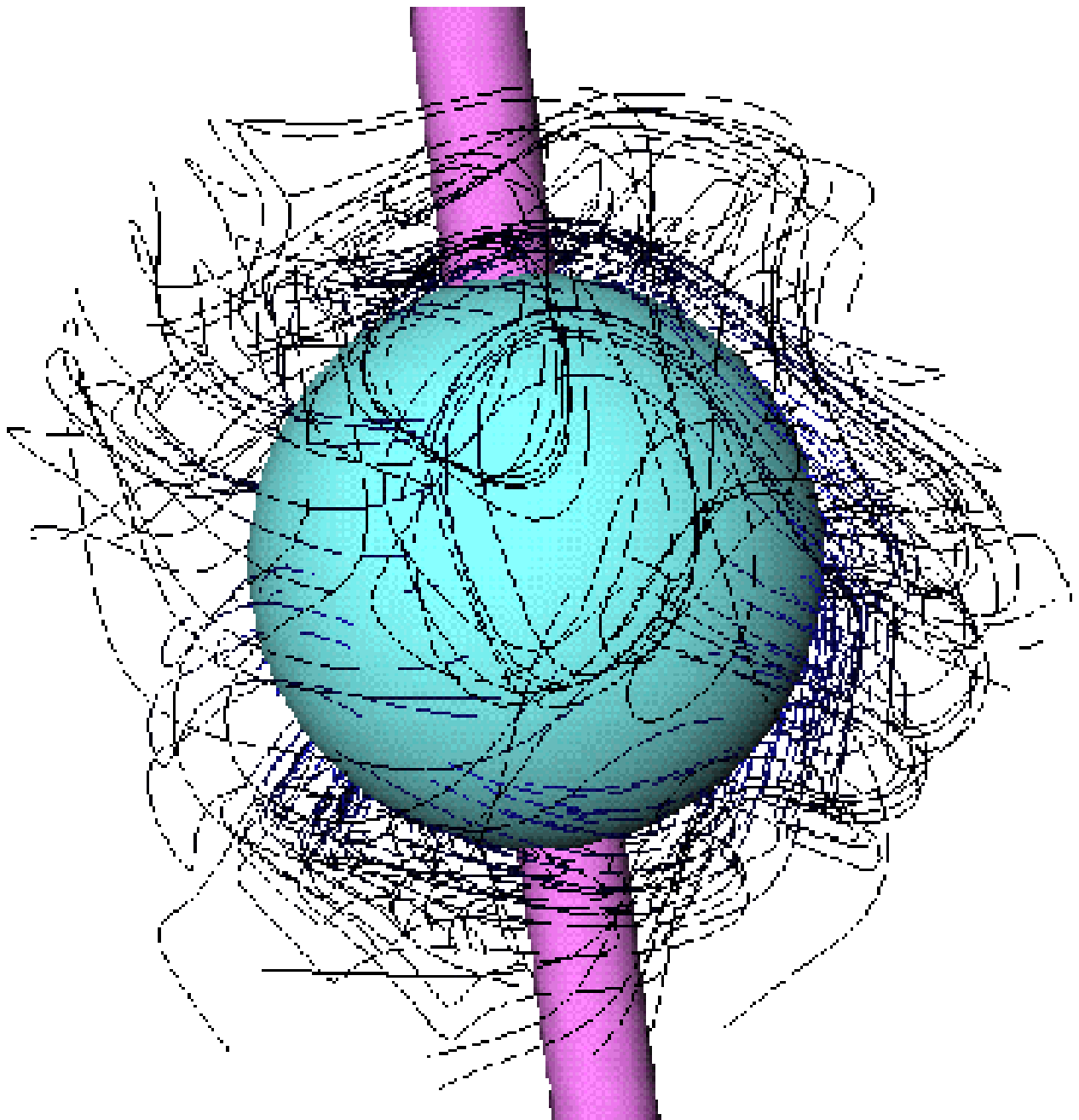}
\includegraphics[width=0.43\hsize]{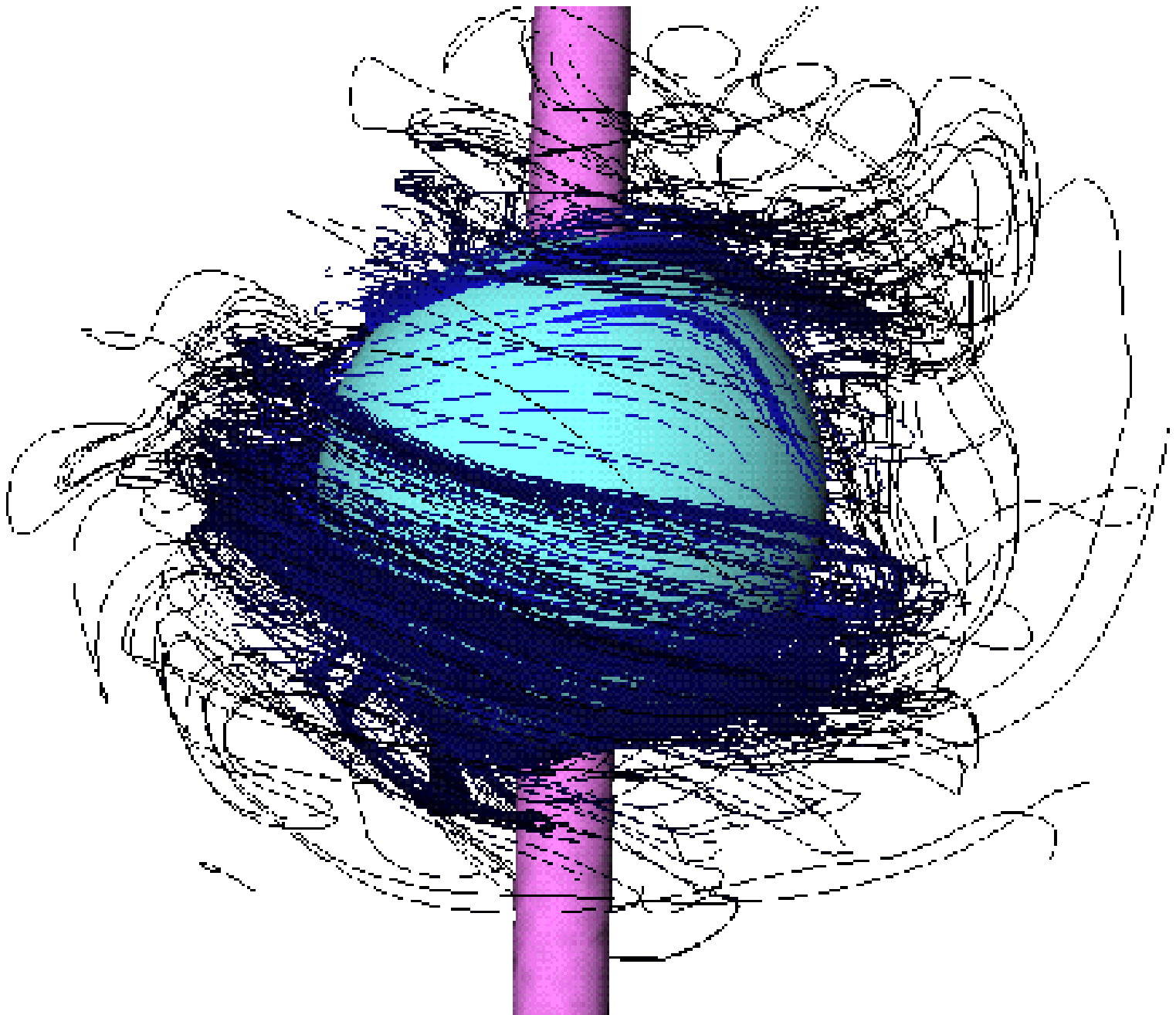}
\includegraphics[width=0.43\hsize]{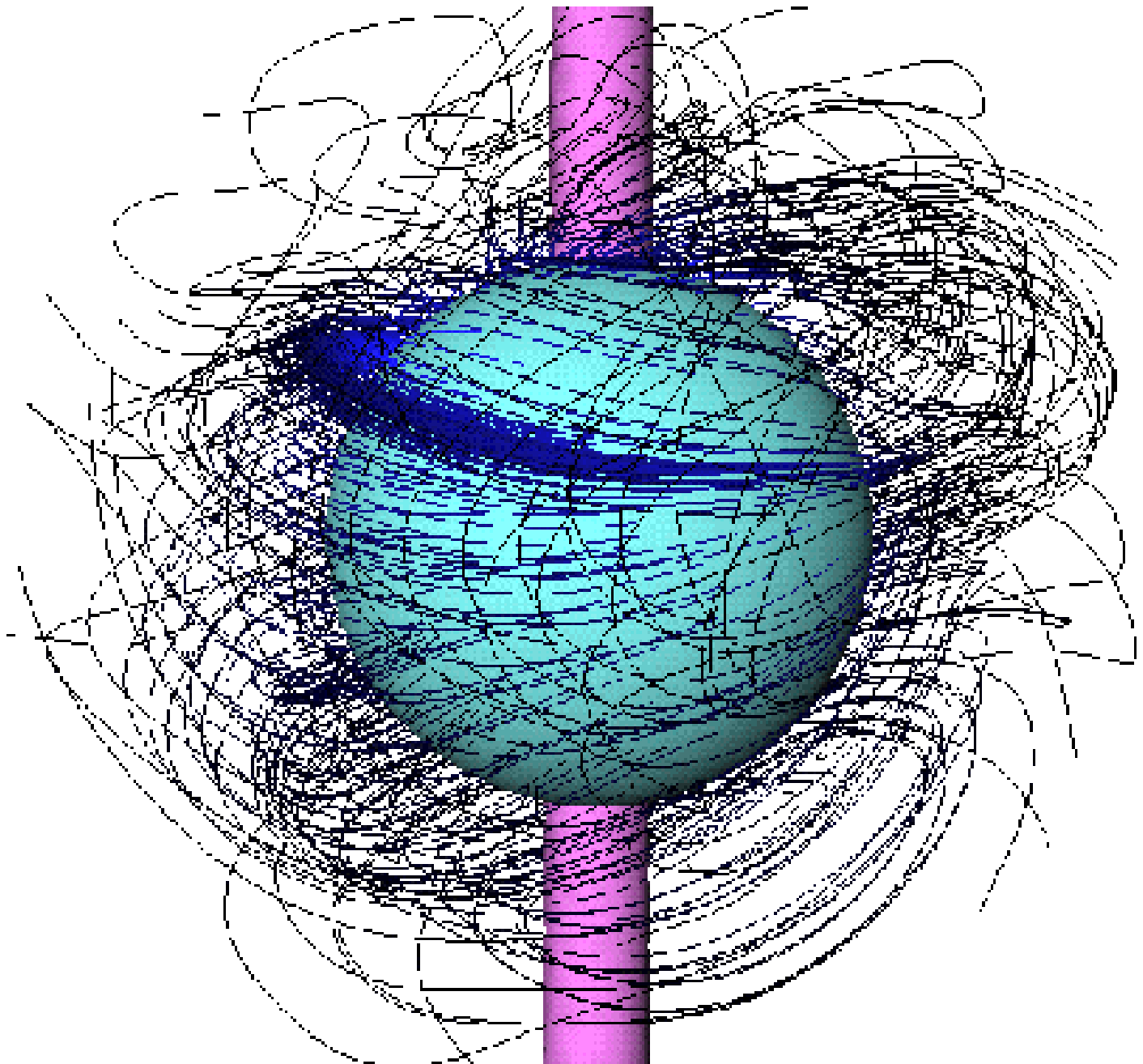}
\includegraphics[width=0.43\hsize]{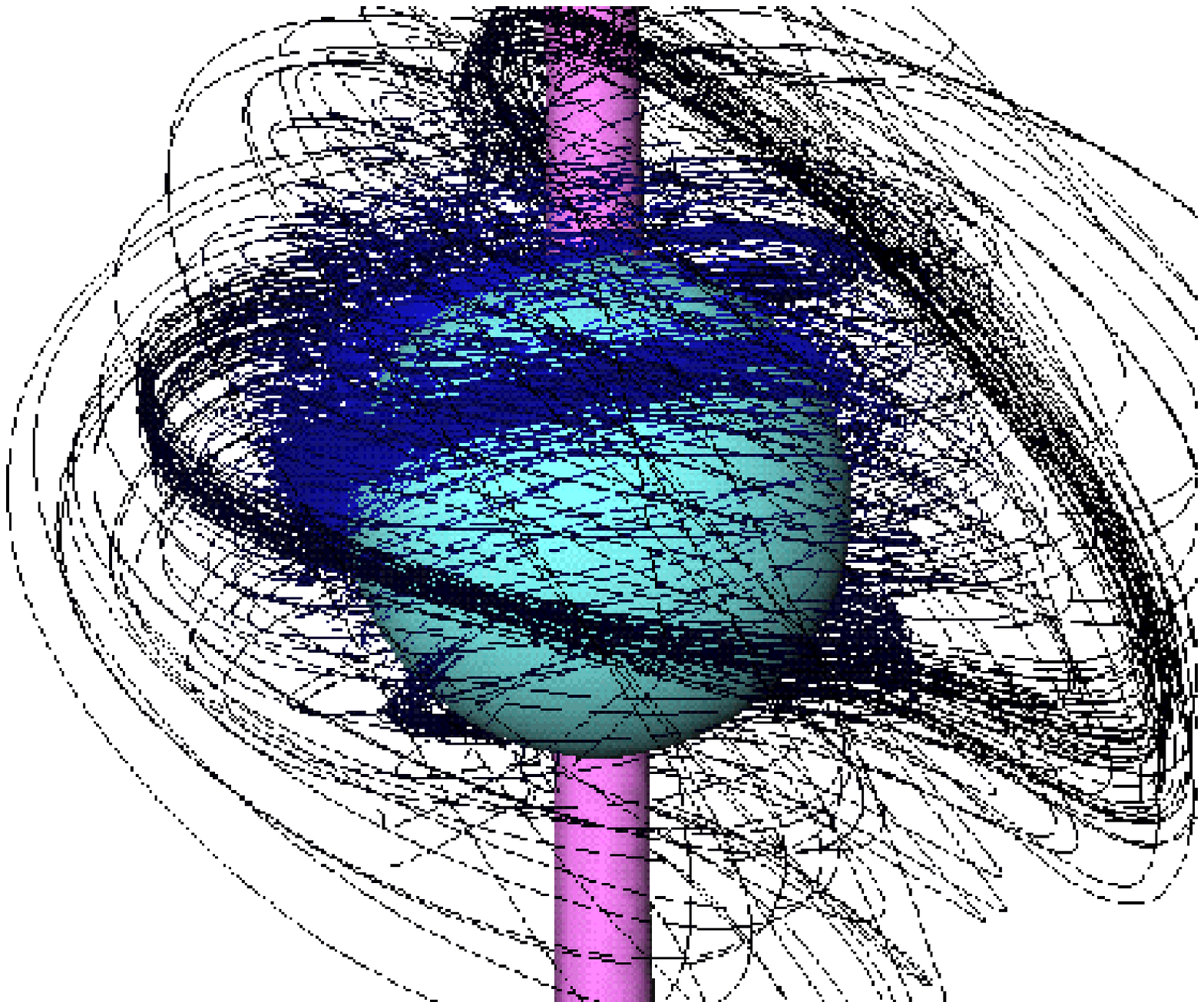}
\hfill
\includegraphics[width=0.43\hsize]{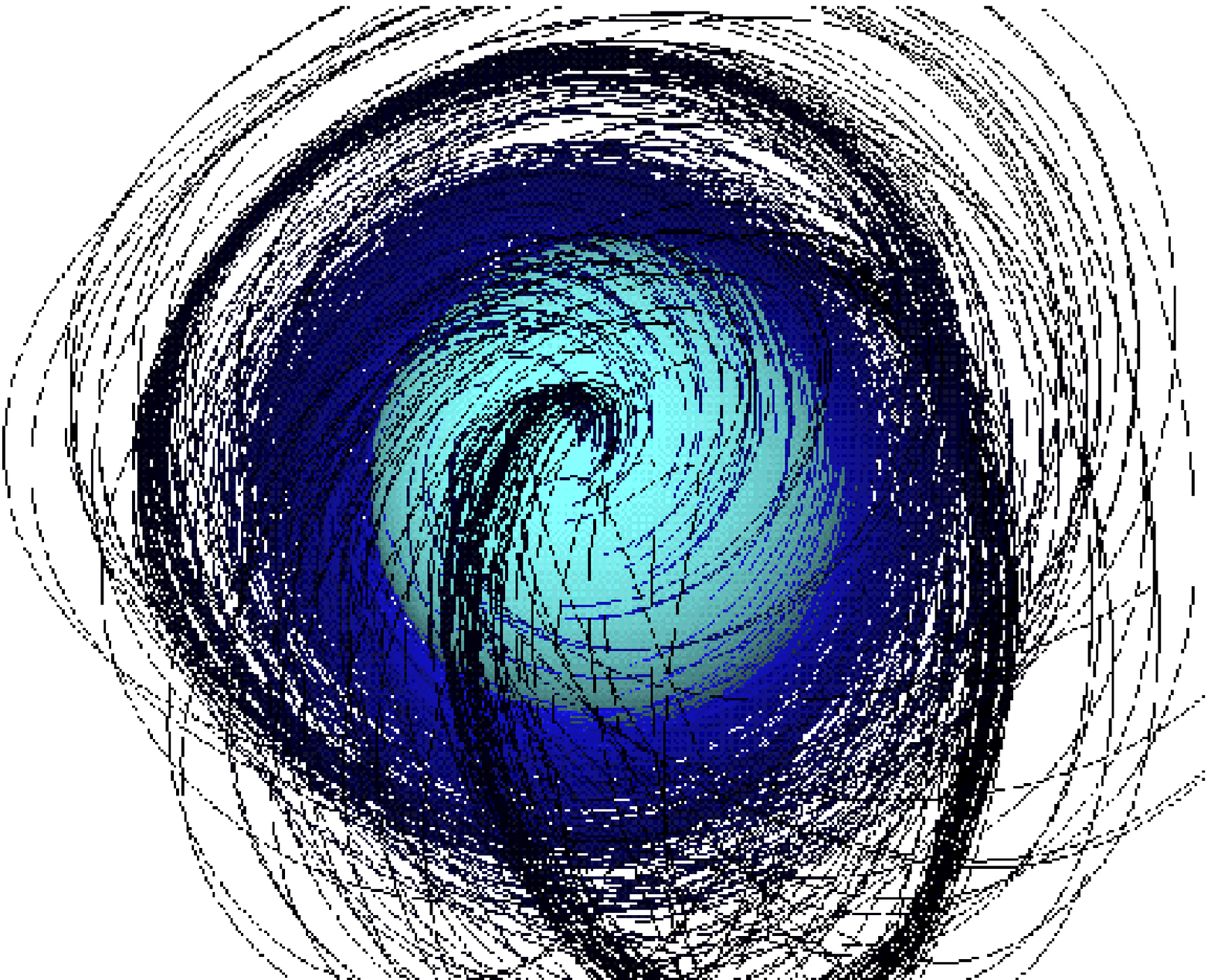}
\caption{The initial evolution of the field, plotted with
\cite{Explorer}. Plotted top left at $t=0$ is the computational box, with field
lines, the axis $\mathbf{M}$
and (only in this frame) the surface of the star. Also plotted in this
frame, as in
all of the frames, is a surface of constant radius ($r=0.3R_\ast$), which helps to make it
easier to see the field lines in the foreground. However, in this
frame this is difficult to see. Therefore, the other frames are
zoomed-in somewhat, so that only the $r=0.3R_\ast$ surface is visible
and not the surface of the star.
Top right is also at $t=0$, but viewed from a different angle, and zoomed-in. Middle
left, middle right and bottom left are snapshots taken at times
$t=0.18$, $0.54$ and $5.4$ days.
Bottom right is the last of these, looking
down the magnetic axis.}
\label{fig:exp-initevol}
\end{figure*}

As described in Sect. \ref{sec:apfa}, relaxation of the magnetic field in 
the atmosphere is an integral part of the stability problem. As a first test, 
however, a field was evolved in a star without atmosphere, at a resolution 
of $96^3$. This evolution would be typical of the evolution of a field at high 
$\beta$, buried inside the star.

After around $3$ Alfv\'en crossing times (based on the initial field strength), 
the field energy has decayed by a factor of around $50$ and has
assumed a configuration which then appears to be stable. The poloidal component is very 
similar to that which would be produced by an azimuthal current loop near the 
equator of the star. The toroidal component then threads along this loop.  The 
loop is generally a little off-centred both in radius and in latitude, and almost 
circular.

This field then gradually diffuses outwards into the atmosphere, maintaining 
its overall form as it does so, until of course it reaches the edges of the 
computational box and the periodic boundary conditions have an effect.

Several cases like this were run, (also at resolution $96^3$), with different 
realisations of the random initial field. The outcome was always similar -- in 
all cases a twisted torus field was produced, either right or left-handed.
In a small proportion of cases, both right and left-handed tori were formed
above one another, in which case one eventually dies away.

This suggests an important first conclusion: there is perhaps only one possible 
type of  stable field configuration in a star. If others exist, they are 
apparently not easily reached from random initial conditions.

Next, we consider cases where the magnetic field is allowed to relax to a
potential field in the atmosphere, by means of the atmospheric diffusion 
term described in Sect. \ref{sec:apfa}. The initial evolution of the field is 
unaffected by the addition of the atmospheric diffusion term, provided that 
the length scale  $r_\mathrm{m}$ of the initial field configuration (cf. Eq. 
(\ref{eq:rm})) is small enough. This comes of course as no surprise, since the
properties of the atmosphere should have no effect on a field confined to 
the stellar interior. Fig. \ref{fig:exp-initevol} shows, in a sequence of snapshots, 
the early evolution of the field, from the initial random state into the
torus shape. For this run, a resolution of $144$ cubed was used --
higher than for the other runs, since a duration of just a few
Alfv\'{e}n-crossing-times was required.
The torus field forms on a timescale of the order of a few Alfv\'{e}n
crossing times, which is equal to around $0.6$ days at this field
strength. The snapshots in the figure are taken at times $t=0, 0.18,
0.54$ and $5.4$ days, i.e. after $0, 0.3, 0.9$ and $9$ Alfv\'en
crossing times. Once the torus is clearly defined, it makes sense to talk 
about toroidal (azimuthal) and poloidal (meridional) components of
the field, by defining them relative to the axis of the torus. The axis 
definition $\mathbf{M}$ (cf. Eq. (\ref{eq:m1})) was used for division 
into toroidal and poloidal components.

By the time the torus field has formed, the field energy has decayed
by a factor of 50 or so. As the field then diffuses gradually
outwards, the effect of the atmospheric diffusion term begins to show
itself. This is because at first, the field is confined to the
interior and consequently unaffected by the properties of the
atmosphere. Once the field has diffused outwards somewhat it will
clearly begin to be affected by the fact that the star has a surface
beyond which the properties of the material are different.

\begin{figure}
\includegraphics[width=1.0\hsize,angle=0]{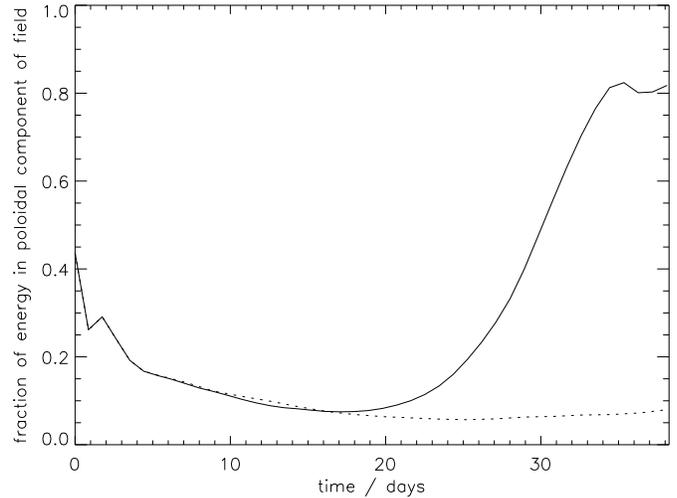}
\caption{The fraction of the magnetic energy contained in the poloidal 
field for the two runs (at resolution $96^3$) with the atmospheric diffusion 
term switched on (solid line) and off (dotted line).}
\label{fig:etaeffect-pc}
\end{figure}

To illustrate this Fig. \ref{fig:etaeffect-pc} shows the energy in the poloidal 
field component, as a fraction of the total magnetic energy (described
in Sect. \ref{sec:vis}). It is seen
that the atmospheric diffusion term causes the poloidal
component of the field to become stronger than the toroidal. When it
is first formed, the torus field has something like 90\% of its energy in
the toroidal component, but a non-conducting atmosphere cannot of
course support a twisted field outside of the star. Only the
poloidal component survives the move from inside to outside. The
energy of the toroidal component therefore falls compared to that of
the poloidal component.

\begin{figure*}
\includegraphics[width=0.5\hsize,angle=0]{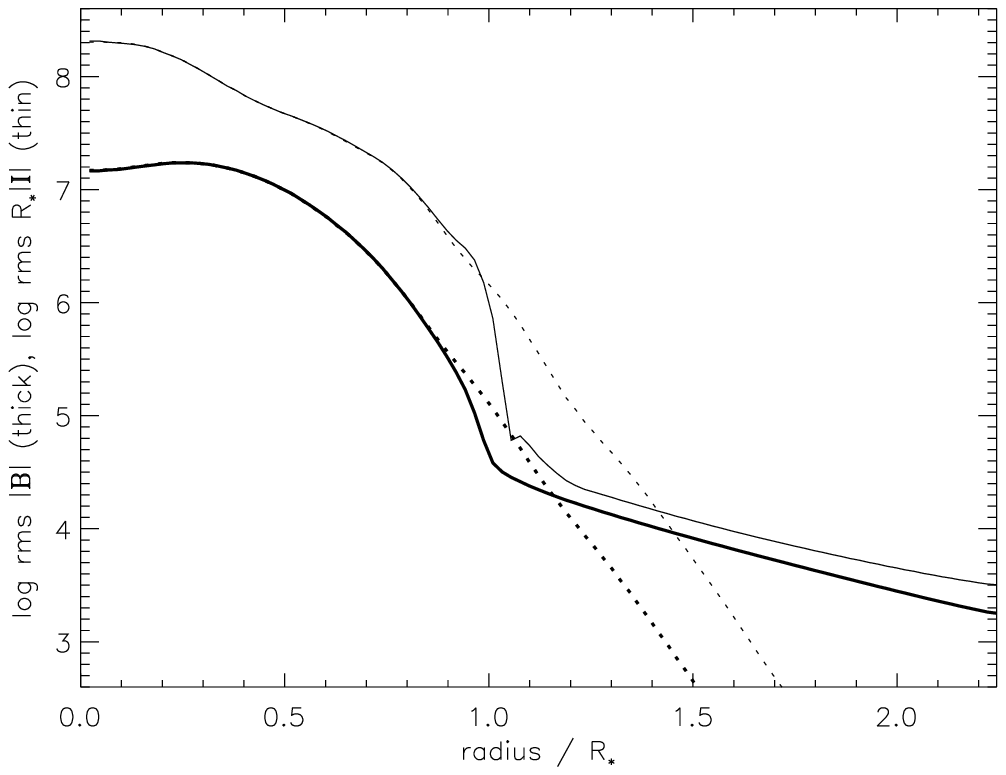}
\includegraphics[width=0.5\hsize,angle=0]{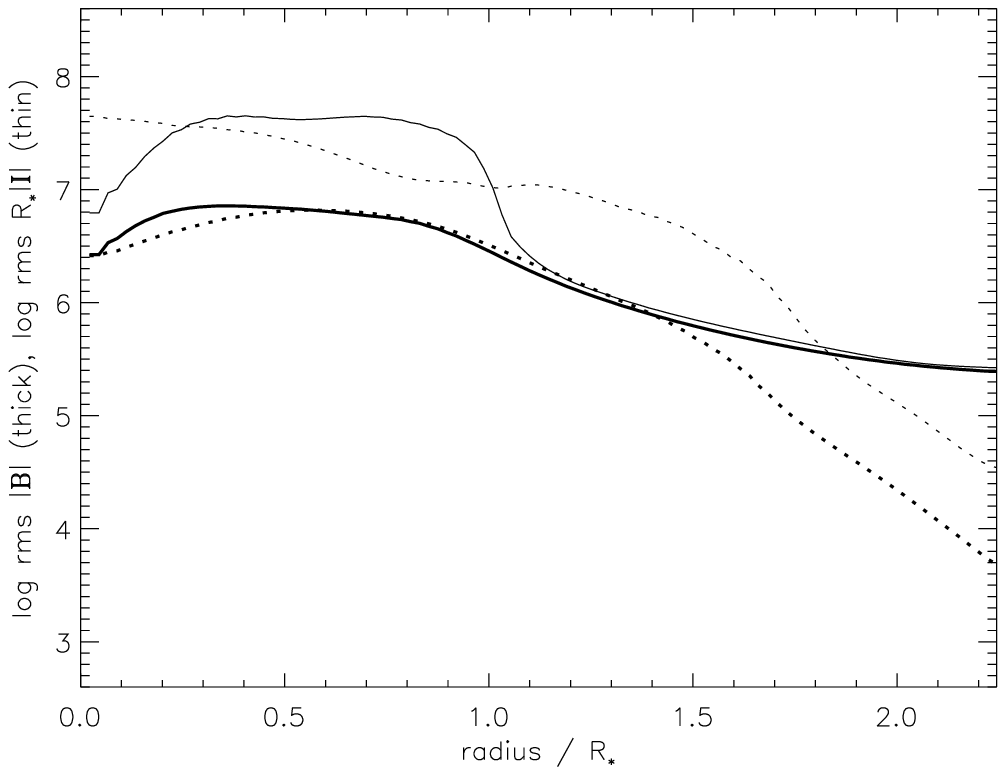}
\caption{Root-mean-square as a function of radius of field amplitude $|\mathbf{B}|$ (thick
lines) and current density $R_\ast|\mathbf{I}|$ (multiplied by $R_\ast$ out of consideration for units)
(thin lines), for the two runs with the atmospheric diffusion
term switched on (solid line) and off (dotted line), at times $t=5.4$
(left) and $t=38.1$ days (right).}
\label{fig:etaeffect-bandri}
\end{figure*}

It is useful to check that this diffusion term is doing its job
properly, i.e. to suppress the electric current in the atmosphere. To
this end, we can look at the current density in the stellar interior
compared to that in the atmosphere. In Fig.
\ref{fig:etaeffect-bandri} we have plotted the radial averages of
the field strength and of the current density (multiplied by the stellar
radius $R_\ast$ to give the same units as field strength), at times
$t=5.4$ and $t=27.2$ days,
i.e. during the slow outwards diffusive phase of the field's
evolution. It is the difference between these two quantities that we
are interested in, and we can see that when the diffusion term is
switched on, the field in the atmosphere is stronger than when it is
switched off -- this is because a potential field, which is what we
have when the term is switched on, responds immediately to the field
on the stellar surface, while a field takes much longer to
penetrate a current-carrying atmosphere. Also, the diffusion term has
the effect of reducing by a factor of ten or so the value of
$R_\ast|\mathbf{I}|$ in relation to the field strength. 
All subsequent discussion is limited to runs performed with the
atmospheric diffusion term switched on.

\begin{figure*}
\includegraphics[width=0.5\hsize,angle=0]{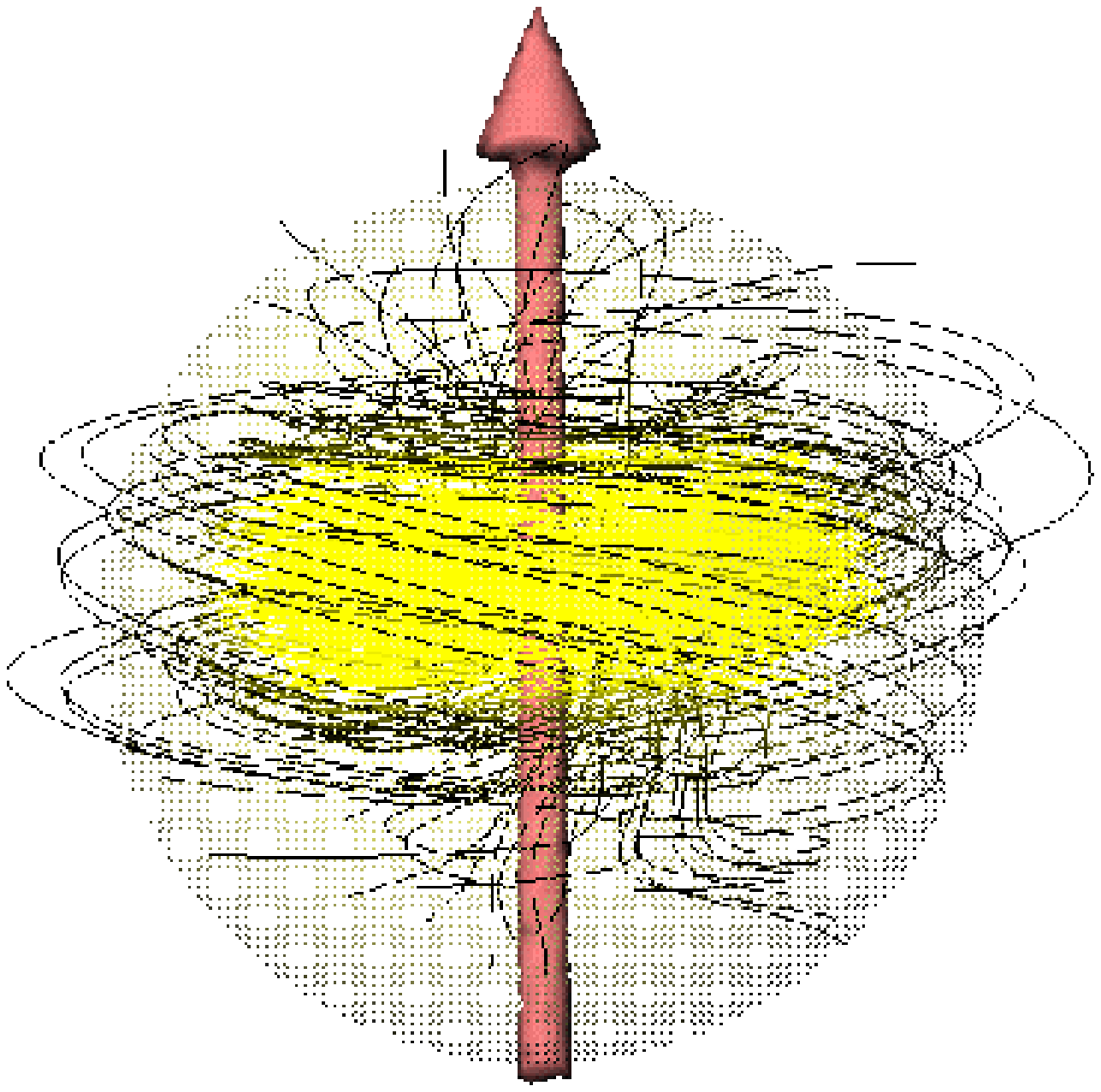}
\includegraphics[width=0.5\hsize,angle=0]{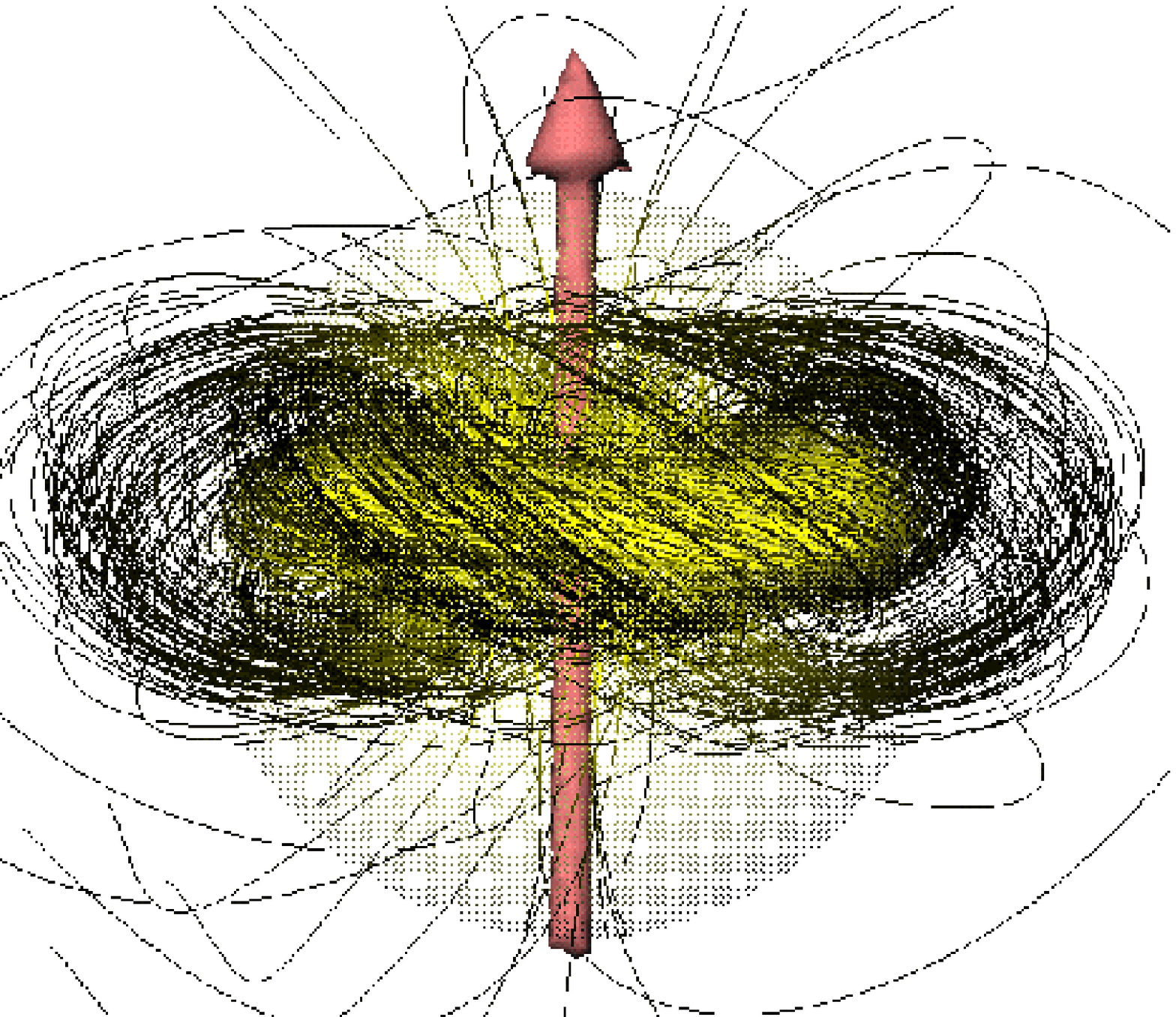}
\includegraphics[width=0.5\hsize,angle=0]{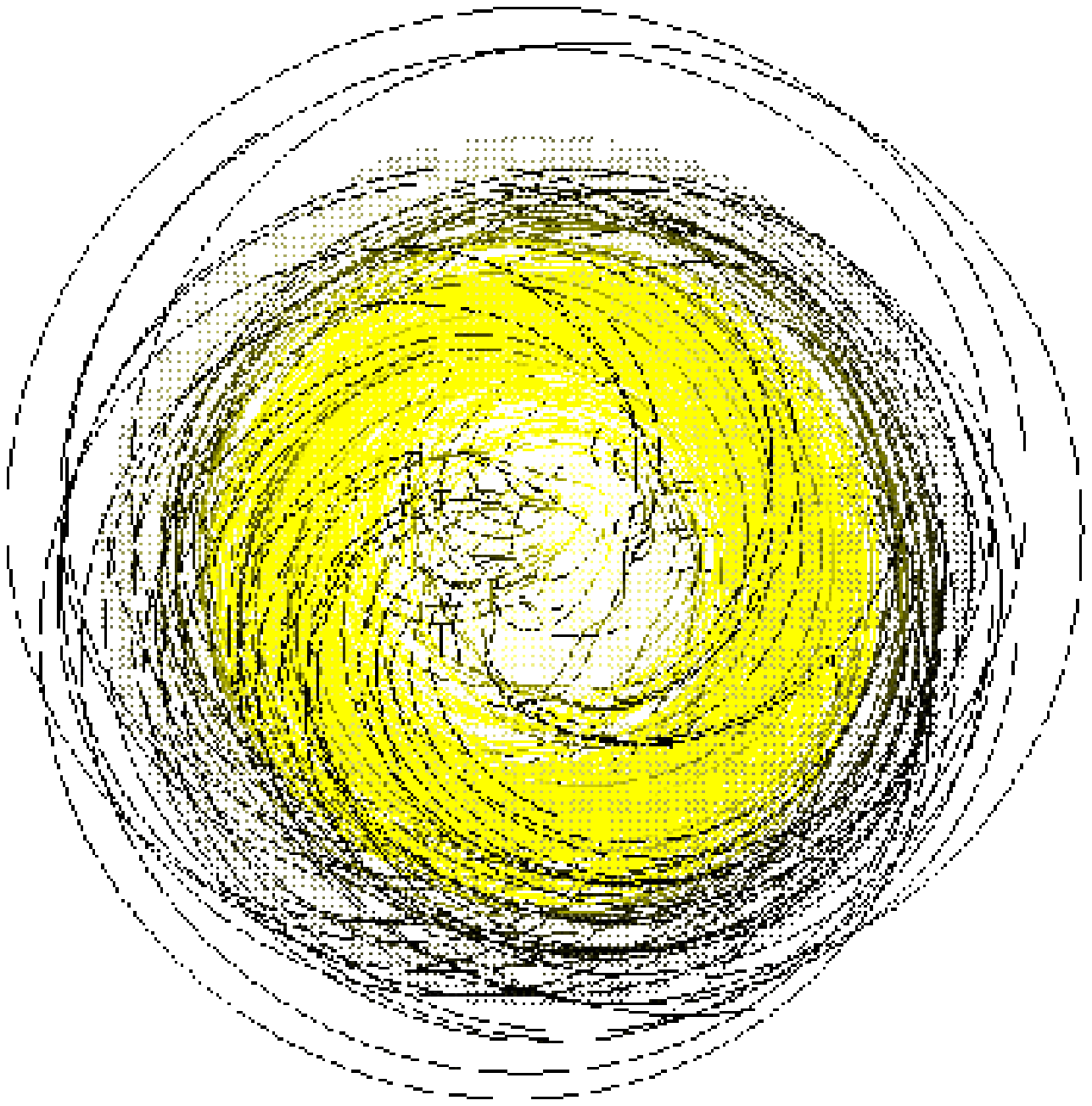}
\includegraphics[width=0.5\hsize,angle=0]{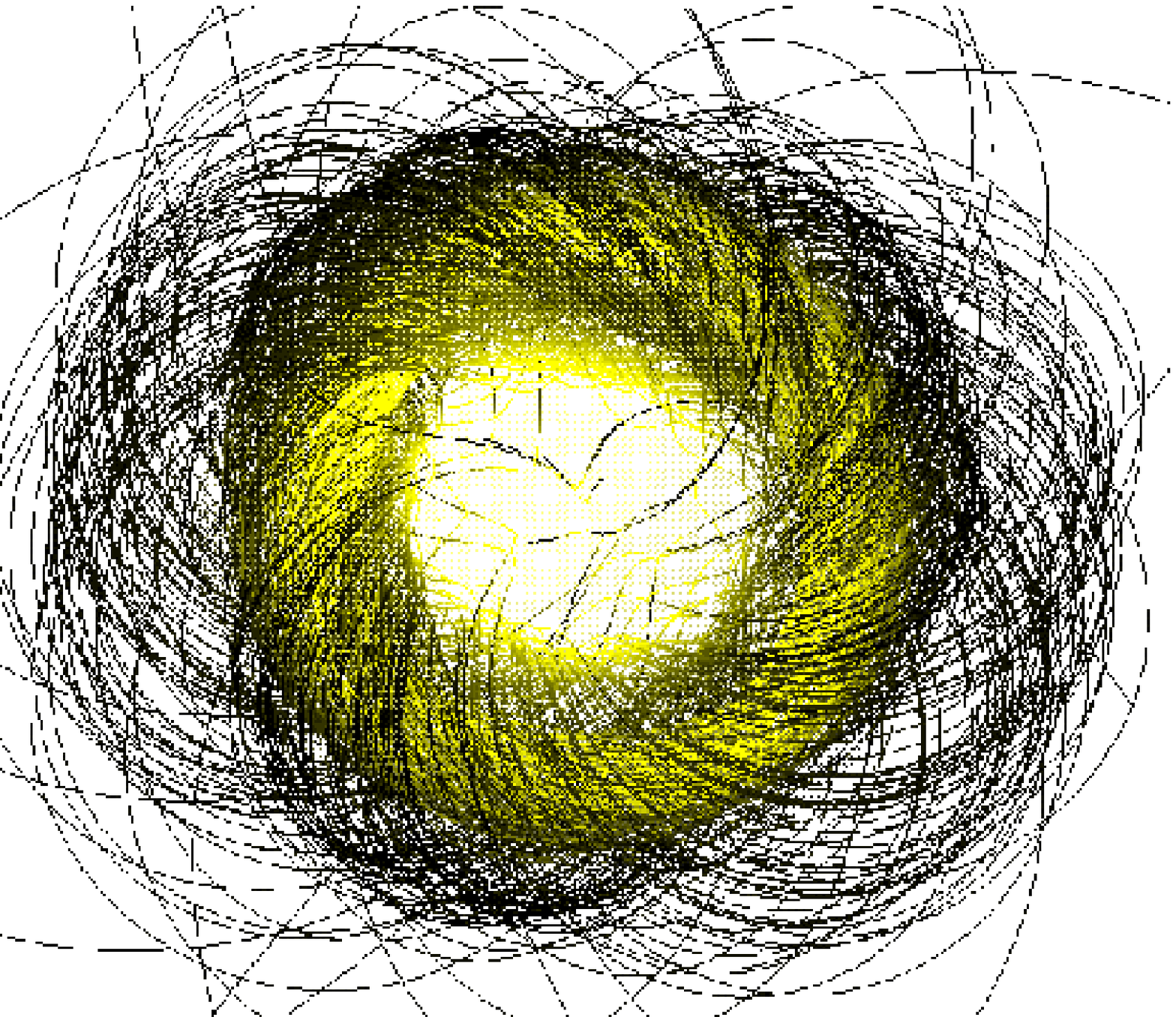}
\caption{Snapshots of the fiducial run at resolution $96^3$. Upper left,
at $t=22.6$ days,
field lines are plotted, in addition to the magnetic
axis $\mathbf{M}$ (the arrow) and
a transparent sphere of radius
$R_\ast$. Lower left, the same viewed from a different angle. Upper right
and lower right, the same at a later time
$t=31.9$ days.
As can be seen here as well as in Fig. \ref{fig:etaeffect-pc}, the proportion of energy in the poloidal field increases significantly between these two times, from $11\%$ to $65\%$. This is caused by the outwards diffusion of the torus field -- a greater proportion of the magnetic energy is in the atmosphere.}
\label{fig:torus}
\end{figure*}

In Fig. \ref{fig:torus}, we can see how the torus changes as it
diffuses outwards. At two times ($t=22.6$ and $31.9$ days) field lines
are plotted -- it is clear that at the time of the first snapshot, the
field is mainly toroidal, but then the poloidal component grows in
relation to the toroidal.

As the field diffuses further outwards, 
the shape of the field changes. The torus starts distorting as if it were a loaded 
spring trapped inside a hollow ball -- it changes first from a circular shape
to the shape of the line on the surface of a tennis ball, and then to
an more contorted shape, as shown in Fig. \ref{fig:foursnaps}.

\subsection{Tests}
\label{sec:test}
The validity and accuracy of the code can be judged from the set of
results
to be published in Braithwaite (2005).
In these calculations, a series of 
stability calculations for toroidal field configurations in stably stratified stars 
are reported and compared with known analytical results. The good 
agreement found there demonstrates the applicability for problems like
the present stellar MHD problem. 

\begin{figure}
\includegraphics[width=1.0\hsize,angle=0]{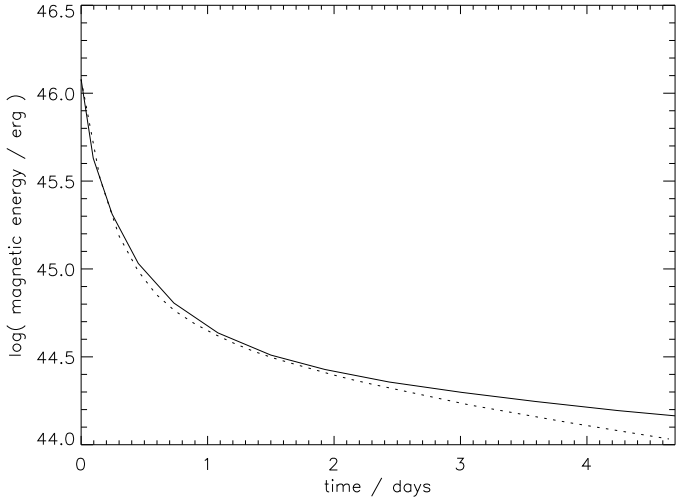}
\caption{\label{fig:uptest-me} Test of the field amplitude rescaling
scheme: The  magnetic energy $E(t)$ is plotted
against the  time $t$, with rescaling (solid line) and without
(dotted line).}
\end{figure}

\begin{figure}
\includegraphics[width=1.0\hsize,angle=0]{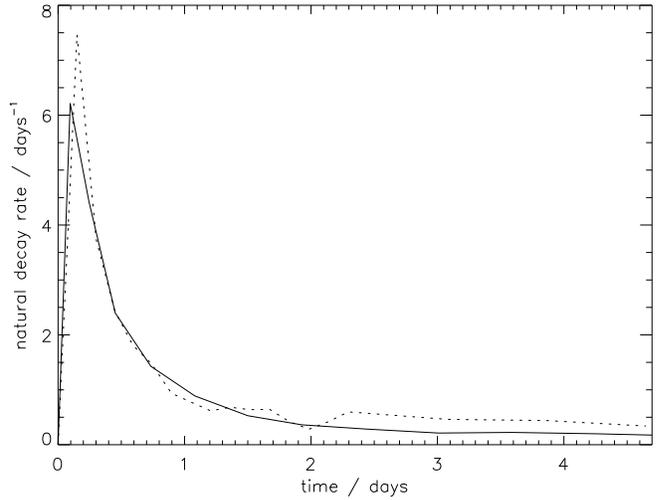}
\caption{\label{fig:uptest-nd}
As Fig. \ref{fig:uptest-me}, but showing the field decay rate $\dot E/E$.}
\end{figure}

As described in Sect. \ref{sec:resca}, a {\it rescaling} procedure is used to 
increase the speed of the calculation. Since this procedure can be formally 
justified only in the limit $\eta/R\ll v_\mathrm{A}\ll c_\mathrm{s}$, 
tests were done comparing the evolution of a given initial field with and 
without this procedure. The result of such a test is shown in Figs. 
\ref{fig:uptest-me} and \ref{fig:uptest-nd}. Plotted, for both runs, are the 
total magnetic energy as a function of time and its decay rate. 
The figures show that the rescaling scheme reproduces field 
decay properly, at least when the field's evolution is primarily on the dynamic 
time scale and not Ohmic. Once the stable field has appeared and diffusive 
processes become important, the scheme ceases to speed up the evolution. This
manifests itself in the two figures in a divergence of the two runs at
later times: the process becomes less accurate when
diffusive processes take over from dynamic evolution. 

\begin{figure*}
\includegraphics[width=0.5\hsize,angle=0]{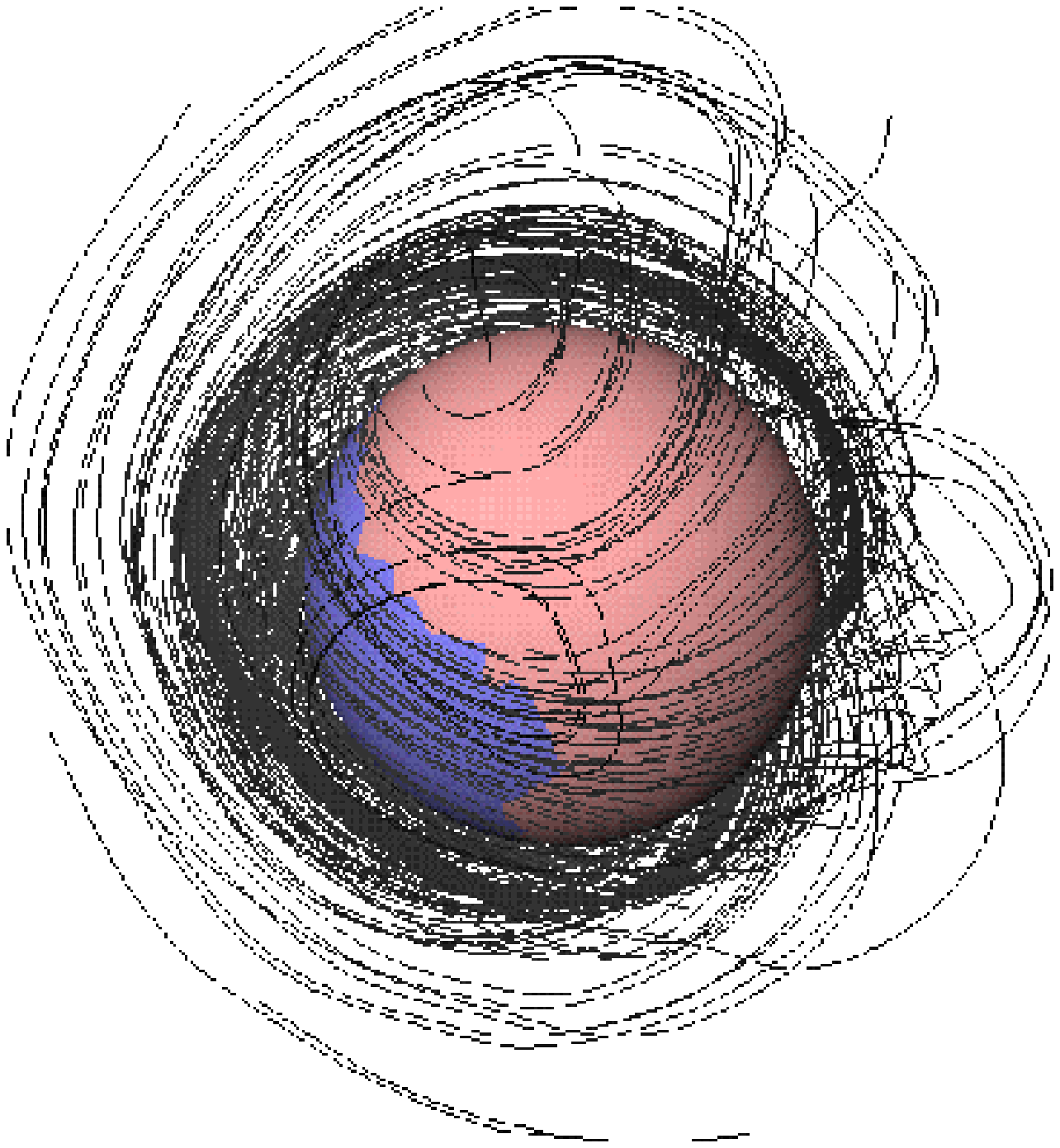}
\includegraphics[width=0.5\hsize,angle=0]{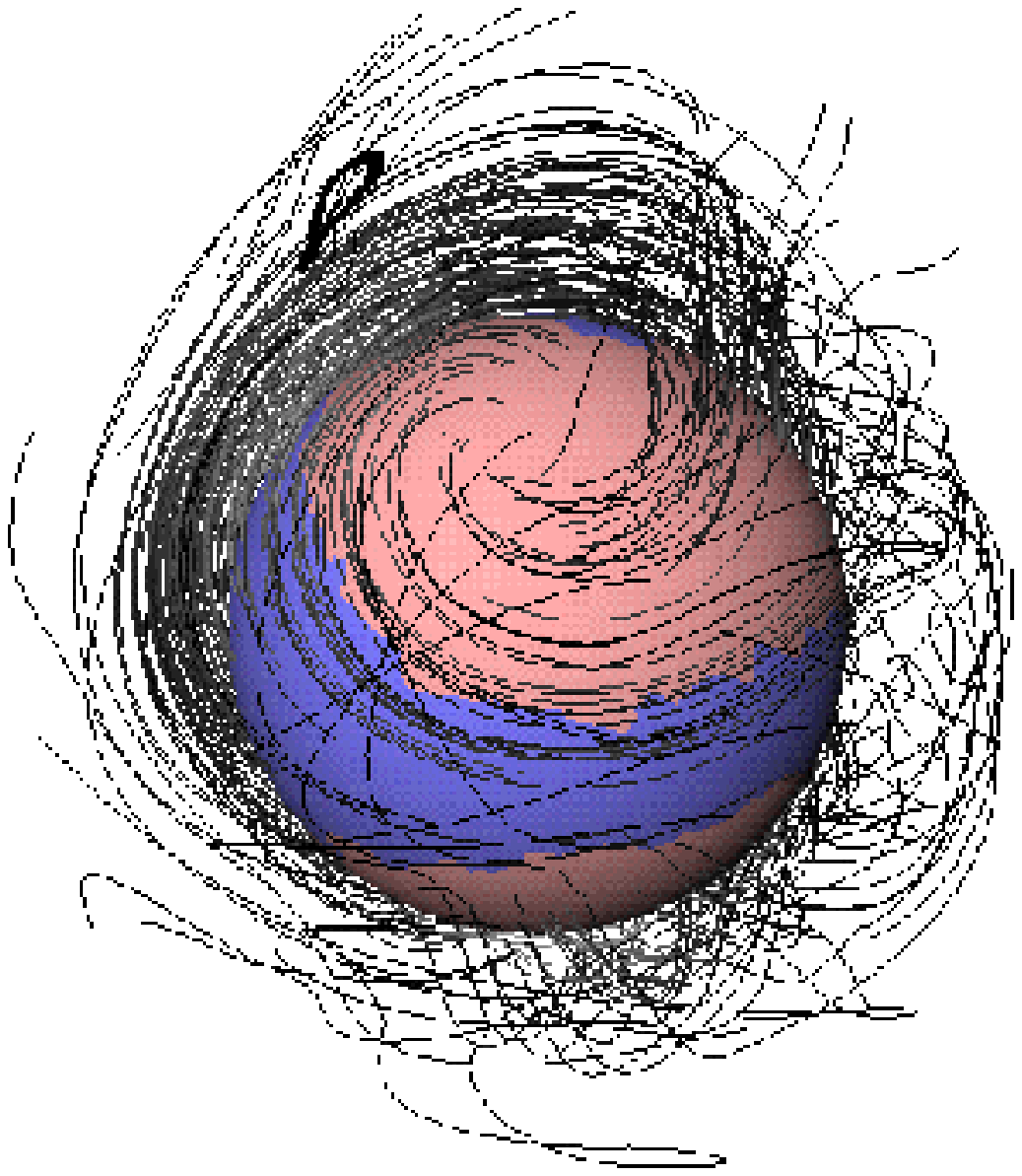}
\caption{\label{fig:uptest-pics} Projections of the field lines without (left) and 
with (right) the field amplitude rescaling scheme (see Sect. \ref{sec:resca}), 
at a common time $t=4.5$ days. Also plotted is a surface
of constant radius $r=0.4R_\ast$, which helps to provide a background
against which to view the field lines.}
\end{figure*}

Fig. \ref{fig:uptest-pics} compares the end result of the evolution of
the field in the two cases at time of $t=4.5$ days. 
This shows the level of difference introduced in the field configuration
by the rescaling process. 


\subsection{The characteristic size of the initial field configuration}
\label{sec:rm}
\begin{figure}
\includegraphics[width=1.0\hsize,angle=0]{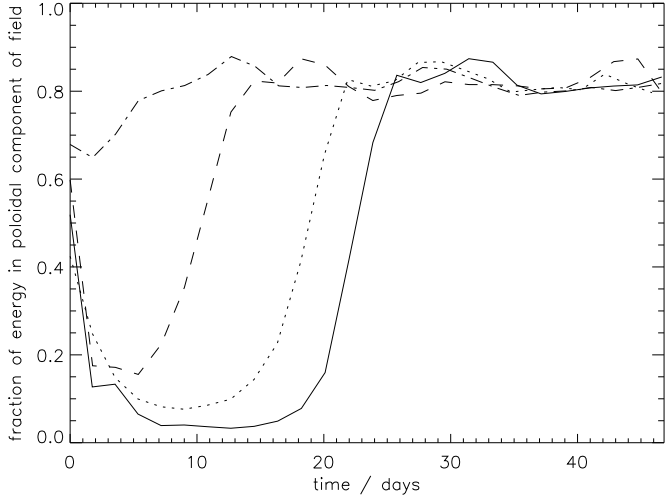}
\caption{The effect of the length scale of the initial field configuration on
the evolution of the field. The poloidal fraction of the magnetic energy is
plotted for $r_\mathrm{m}=0.14R_\ast$ (solid line), $0.25R_\ast$ (dotted), $0.39R_\ast$
(dashed) and $0.57R_\ast$ (dot-dashed). It can be seen that the initial
conditions merely determine the route taken to the final state, not
the final state itself.}
\label{fig:rmeffect-pc}
\end{figure}

The evolution is found to depend on the initial state of the magnetic field, or 
at least, on the initial length scale $r_\mathrm{m}$
(cf. Eq. (\ref{eq:rm})) of the field. 
For small $r_\mathrm{m}$, the field configuration is concentrated more
towards the centre of the star. Runs were done with different values of 
$r_\mathrm{m}$ but with the same total magnetic energy. 
The field finds the torus configuration only if $r_\mathrm{m}$ is below a certain
value, so that if $r_\mathrm{m}$ is smaller than this value, the torus produced
diffuses gradually outwards until at some point it  starts the `tennis ball' 
distortion described above. The smaller $r_\mathrm{m}$ is, the smaller the 
torus produced is, and the longer this  diffusive phase lasts. If $r_\mathrm{m}$ 
is above the critical value, the field goes straight into the distorted state without 
ever reaching the regular torus shape. So its value has no effect on the final 
state of the field, i.e. one of fast decay caused by dynamic instability. In the 
runs described above, we used $r_\mathrm{m}=0.25R_\ast$. To
look at the effect of $r_\mathrm{m}$, we did otherwise identical runs
at resolution $72^3$ with $r_\mathrm{m}=0.14R_\ast$, $0.25R_\ast$,
$0.39R_\ast$ and $0.57R_\ast$. These runs can be compared by looking at the
poloidal field energy as a proportion of the total energy -- see Fig.
\ref{fig:rmeffect-pc}. It can be seen that the value of $r_\mathrm{m}$
affects the route taken to reach the final state, but has no effect on
the final state itself. The run with $r_\mathrm{m}=0.57R_\ast$ never
reaches any stable state at all; the other three runs with smaller
$r_\mathrm{m}$ do reach it, and stay in it longer the smaller 
$r_\mathrm{m}$ is.

\begin{figure}
\includegraphics[width=1.0\hsize,angle=0]{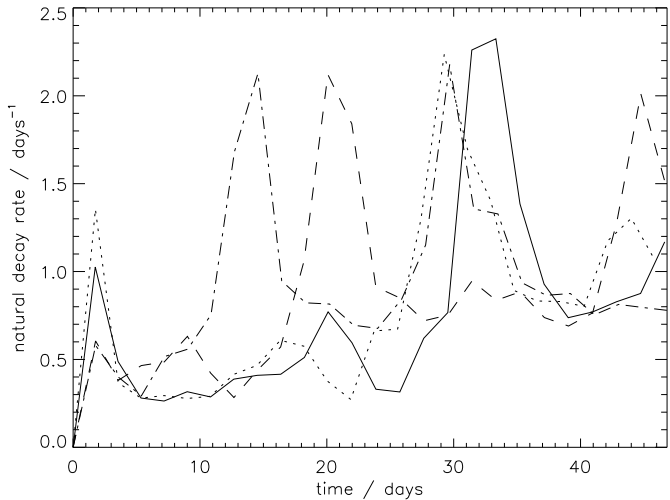}
\caption{The rate of decay of the field, for the four runs with different values of
$r_\mathrm{m}$.}
\label{fig:rmeffect-nd}
\end{figure}

This can be confirmed by looking at the rate of decay of the field, as plotted 
in Fig. \ref{fig:rmeffect-nd}. It can be seen that the greater $r_\mathrm{m}$, 
the sooner the point is reached at which the field begins to decay quickly.


\subsection{A quantitative look at the diffusive phase of
evolution}
\label{sec:phase2}

Once the stable torus field has formed, it gradually diffuses
outwards. If the configuration was initially concentrated towards the
centre, it is then possible that the strength of the field
on the surface increases, despite the fact that the total magnetic
energy goes down. 

To look at this in a quantitative manner, we have introduced a diffusivity 
with the functional form of the Spitzer's (1962) conductivity for ionised
plasmas, applicable in stellar interiors:
\begin{equation}
\eta_0=K T^{-3/2}.
\label{eq:physdiff}
\end{equation}
Here $K\approx 7\times10^{11}\ln \Lambda$ where the Coulomb 
logarithm $\ln\Lambda$  is of order $10$ in a stellar interior.

Adding this diffusivity to the code would result in the field evolving 
much too slowly to be computationally practical. We can make use of 
the fact that, in the stage we are interested in here, the field is evolving 
on the diffusive time-scale. In this case, an increase of the diffusivity by a 
constant factor, while maintaining the functional dependence (\ref{eq:physdiff}), 
is equivalent to a decrease in the time-scale of evolution. Thus, we use
a diffusivity of the form (\ref{eq:physdiff}), with $K$ adjusted to yield
a speed of evolution that is sufficiently long compared with the Alfv\'en
time-scale, but still short enough to be computationally feasible. The 
evolution can then be scaled afterwards to a realistic time axis.

As the initial conditions, we used
the output from the fiducial run at resolution $96^3$ at a time $t=4.4$
days -- once the stable torus field has formed. The numerical
diffusion scheme, which is required to hold the code stable when the
field is evolving on a dynamic time-scale, was switched off for these
runs. The field rescaling routine (Sect. \ref{sec:resca}) was also switched off.
We ran the code with $\eta/\eta_0$ equal to $10^{11}$, $1.7\times10^{11}$,
$3\times10^{11}$, $5.5\times10^{11}$, $10^{12}$, $1.7\times10^{12}$
and $3\times10^{12}$.

\begin{figure}
\includegraphics[width=1.0\hsize,angle=0]{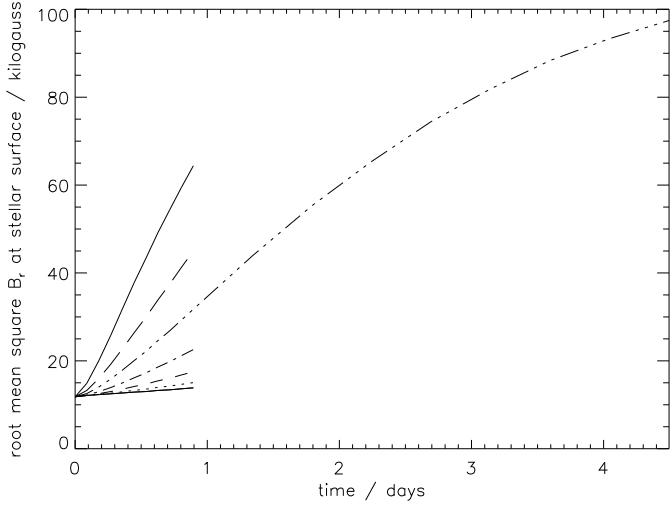}
\caption{Root-mean-square $B_r$ at the surface of the star, as a
function of time. Seven values of diffusion: $10^{11}\eta_0$ (solid),
$1.7\times10^{11}\eta_0$ (dotted), $3\times10^{11}\eta_0$
(dashed), $5.5\times10^{11}\eta_0$ (dot-dashed), $10^{12}\eta_0$
(dot-dot-dot-dashed), $1.7\times10^{12}\eta_0$ (long-dashed) and
$3\times10^{12}\eta_0$ (solid). The
higher the diffusion, the faster the field on the stellar surface increases.}
\label{fig:phase2-nfu}
\end{figure}

We are interested in what happens to the field strength on the surface
of the star during this phase of evolution, since only the surface
field is observed. Fig. \ref{fig:phase2-nfu} is a plot of this surface
field (to be precise, the root-mean-square of its modulus) as a
function of time, for the runs with different values of
$\eta/\eta_0$. The field strength is indeed found to increase; the
higher the diffusivity, the faster the surface field grows.

Looking at the result of Hubrig et al. 2000a, which suggests that Ap
stars typically become visibly magnetic after 30\% of their 
main-sequence lifetime (which works out at around $3\times10^8$ years),
it would be interesting to see how quickly the surface field in these
runs is rising. We can obtain a time-scale if we divide the field
strength by its time derivative. If we do this for the
$\eta/\eta_0=10^{12}$ case, we obtain the time-scale $0.8$ days; we can
therefore infer that if we set $\eta=\eta_0$, we would measure a
timescale $2\times10^9$ years. 
This is somewhat larger than the main sequence lifespan, but still within 
an order of magnitude.

We conclude that Ohmic diffusion of an internal magnetic field is a plausible 
model for the increase of the surface magnetic field with time implied by 
the observations of Hubrig et al. Quantitative improvements in the physics 
used (stellar structure model, precise value of $\eta$) and numerical resolution
will be needed, however, to test this idea more securely.  

\begin{figure}
\includegraphics[width=1.0\hsize,angle=0]{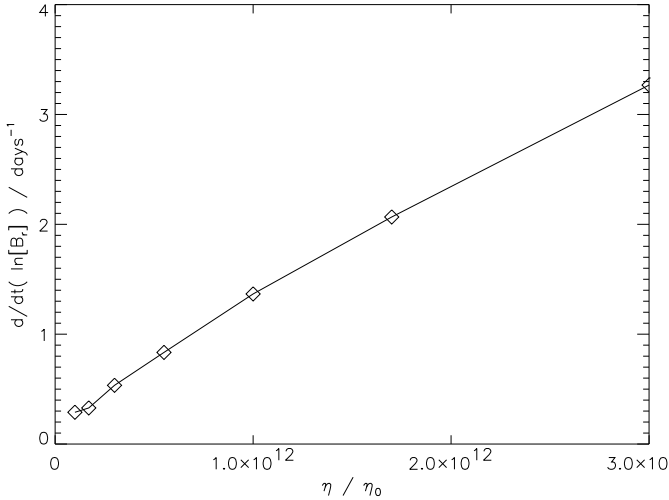}
\caption{The reciprocal of the time-scale on which the surface field
is increasing, $\bar{B_r}/\partial_t \bar{B_r}$, as a function of $\eta/\eta_0$.}
\label{fig:phase2-etatsr}
\end{figure}

It is useful to check that the time-scale really is dependent on the
diffusivity in the way we have assumed, i.e. that the two are
inversely proportional. To this end, we have plotted the reciprocal of
the timescale measured as a function of $\eta/\eta_0$ in
Fig.~\ref{fig:phase2-etatsr}. The two are found to be proportional to
each other, except at the two ends of the range where other numerical
effects come into play.

\begin{figure*}
\includegraphics[width=0.5\hsize]{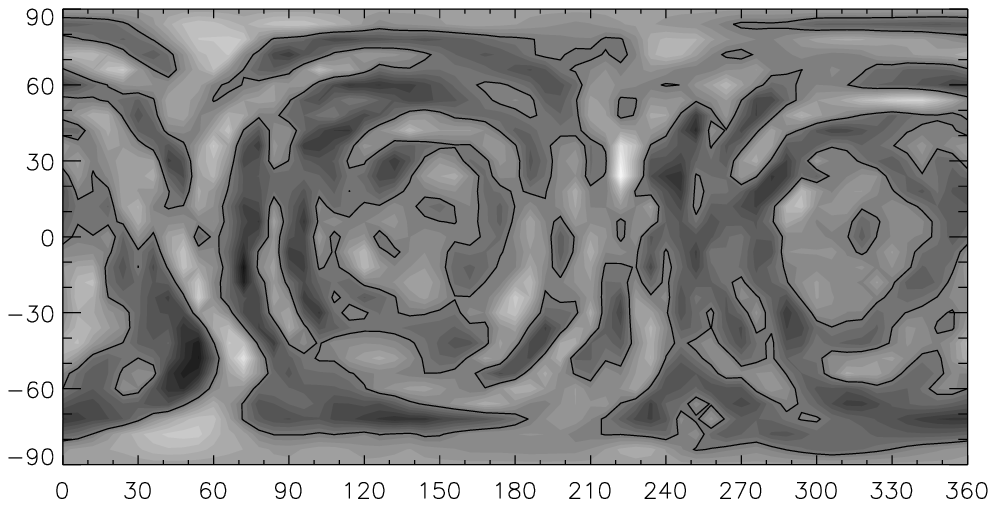}
\includegraphics[width=0.5\hsize]{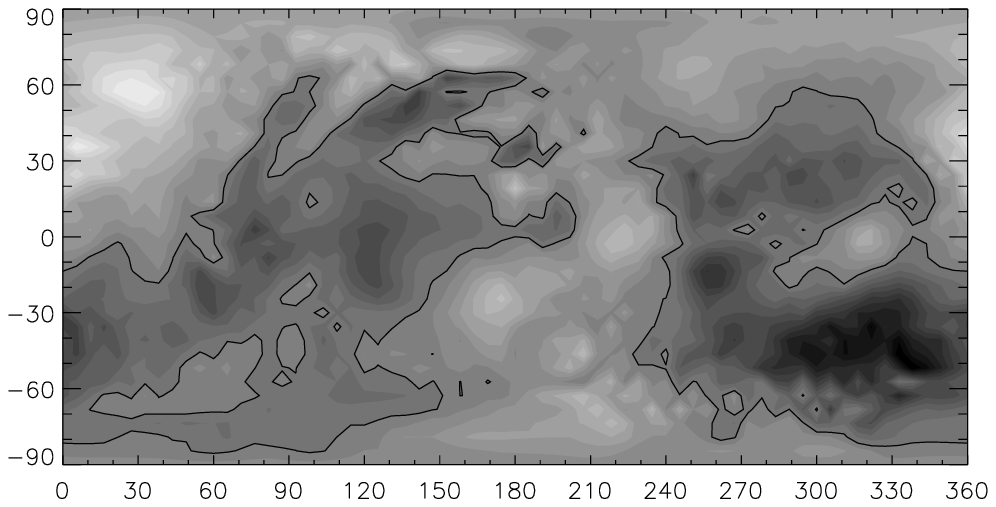}
\includegraphics[width=0.5\hsize]{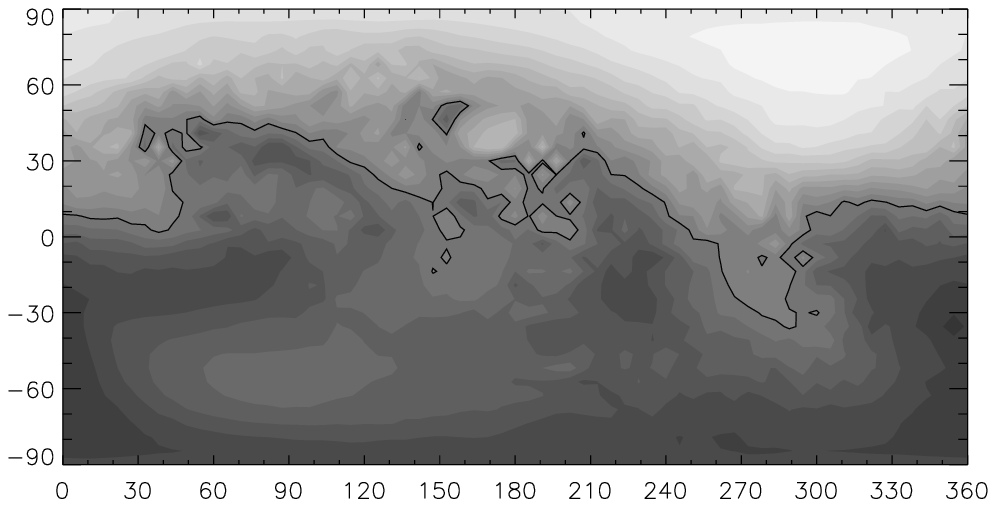}
\includegraphics[width=0.5\hsize]{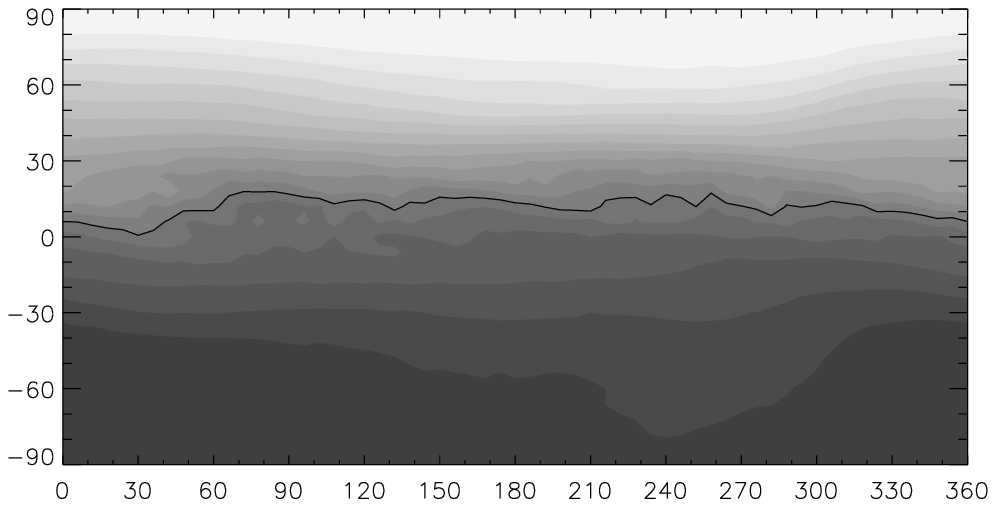}
\caption{Projections onto 2-D of the radial component $B_r$ on the
stellar surface, for the fiducial run at resolution $96^3$, at times
$t=0, 2.6, 9.8, 22.6$ days, using the
axis $\mathbf{M}$. The plots are arranged in the following order:
top-left, top-right, upper-middle-left, upper-middle-right, etc.}
\label{fig:2Dprojectphase2}
\end{figure*}

\subsection{Change in the surface field shape during the diffusive phase}
\label{sec:phase2shape}

We have seen in the last section that the strength of the field on the
surface of the star goes up during the diffusive phase of evolution,
after the stable torus field has formed in the interior. However, this
is not the only change visible to an outside observer -- the {\it shape} of
the surface field also changes. Immediately after the stable field
configuration has been reached in the interior, the field in the outer
part of the star has yet to relax to a dipolar shape; which does then
gradually happen, but only on the longer Ohmic time-scale. This
implies that a field configuration consisting of a regular torus field
in the centre of the star and an irregular field in the outer part can
be dynamically stable.

To illustrate this, it is first helpful to look back at
Figs.~\ref{fig:uptest-nd} and \ref{fig:uptest-me}, in which we can see
that the stable field configuration has formed within one day or so,
i.e. on the order of the Alfv\'{e}n crossing-time, and that after
that the field evolves by mainly diffusive processes, on an Ohmic
time-scale.

Fig.~\ref{fig:2Dprojectphase2} is a projection onto two dimensions of
the field's radial component $B_r$ on the stellar surface.
%
This is plotted at four points in time, the first being $t=0$, the
second just after the stable field has formed at $t=2.6$ days, and the
third and fourth at $t=9.8$ and $22.6$ days respectively. It is easy to
see how the field on the surface approaches a dipole.

\begin{figure}
\includegraphics[width=1.0\hsize,angle=0]{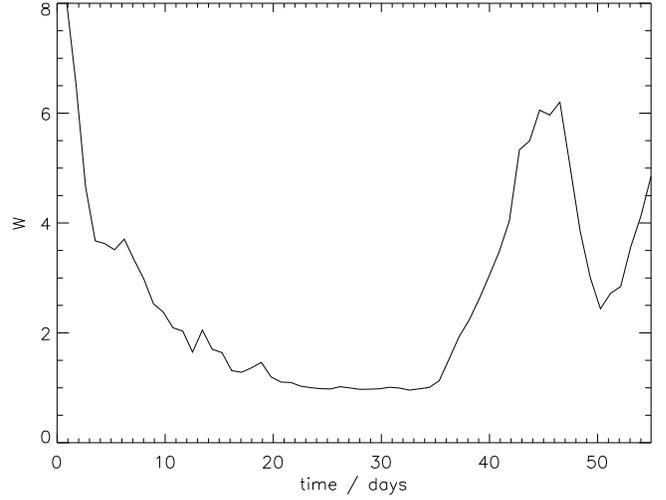}
\caption{$W$, plotted above, is defined as the
length of the line separating positive from negative
$B_r$ on the surface of the star, divided by $2\pi R_\ast$. During the
diffusive phase, this has its dipole value of unity; when the field
becomes unstable it grows.}
\label{fig:wiggliness}
\end{figure}

This can also be looked at in a more quantitative manner. To this end,
we need to find a way to define a length scale on the stellar surface
-- we use the quantity $W$ defined in Sect.~\ref{sec:vis} as the total
length of the line(s) on the surface of the star which separate(s)
regions of positive and negative $B_r$, divided by $2\pi R_\ast$.
(We may then expect that the typical length scale $\mathcal{L}$ is
given by the area of the surface of the star divided by the length of
this $B_r=0$ line, so that $\mathcal{L}\sim 2R_\ast/W$.)
The quantity $W$ in the fiducial run at resolution 96 is plotted in
Fig.~\ref{fig:wiggliness}. It begins large, slowly falls to its
minimum value (of unity) as the field in the outer part of the star
relaxes to a regular dipolar shape and as the torus field diffuses
outwards, and then (as described in the next section) suddenly grows.

\begin{figure}
\includegraphics[width=1.0\hsize,angle=0]{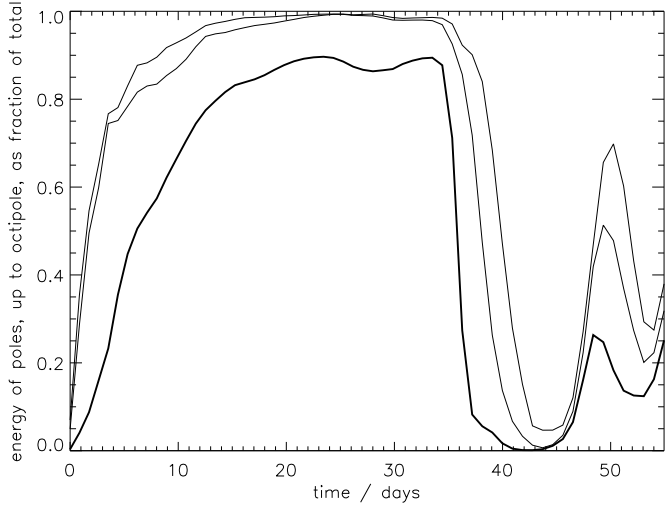}
\caption{The energy of the $B_r$ component on the surface of the star,
broken down into its dipolar, quadrupolar, octupolar and higher
components, as proportions of the total energy. The dipolar
energy is represented by the space between the x axis and the thick
line, the quadrupolar energy is represented by the space between the
thick line and the one above it, etc. The transition from stable to
unstable at $t\approx 35$ days can be seen, as the dipole component
suddenly loses its energy and the surface field becomes dominated by higher
components, first by quadrupole and octupole, then by even higher orders.}
\label{fig:sh-comps}
\end{figure}

It is also interesting to look at the decomposition into spherical
harmonics of $B_r$ on the stellar surface. The proportions of energy
in each spherical harmonic order is plotted in
Fig.~\ref{fig:sh-comps}. When the stable torus field has just formed,
most of the energy is in the higher-order components, and as the
field gradually diffuses outwards, almost all of the energy goes
into the dipole component, with the quadrupolar component making up
almost all of the rest (remember that an offset dipole can be
expressed as dipole plus quadrupole). At $t=22.6$ days, corresponding
to the last frame of Fig.~\ref{fig:2Dprojectphase2}, virtually no
energy is present in the octupole or higher orders. As described in
the next section, the field then later becomes unstable, and the
energy goes back into the higher components.

From this, it is possible to make a hypothesis: that Ap stars with
near-exact dipolar fields are likely to be older than Ap stars with
more structure on smaller scales. However, there will factors other
than age which we would expect to determine how dipole-like the
surface field is, for instance, the degree to which the field was
concentrated into the centre of the star at the time of formation -- a
highly concentrated initial field would result, after formation of the
stable torus field, in a field at the surface with significant
higher-order components. The relative importance of these factors is
uncertain.

\subsection{The final, unstable phase of evolution}
\label{sec:phase3}
As mentioned above, when the stable torus field diffuses outwards to a
certain radius, it eventually becomes unstable and decays. The shape of the 
field changes from an ordered, large-scale shape to a disordered,
small-scale shape which then constantly changes and moves around. This
fall of length scale brings about an increase in the rate at
which energy is lost via Ohmic diffusion, since the time-scale over
which the latter occurs is proportional to the square of the length
scale.

\begin{figure*}
\includegraphics[width=0.48\hsize]{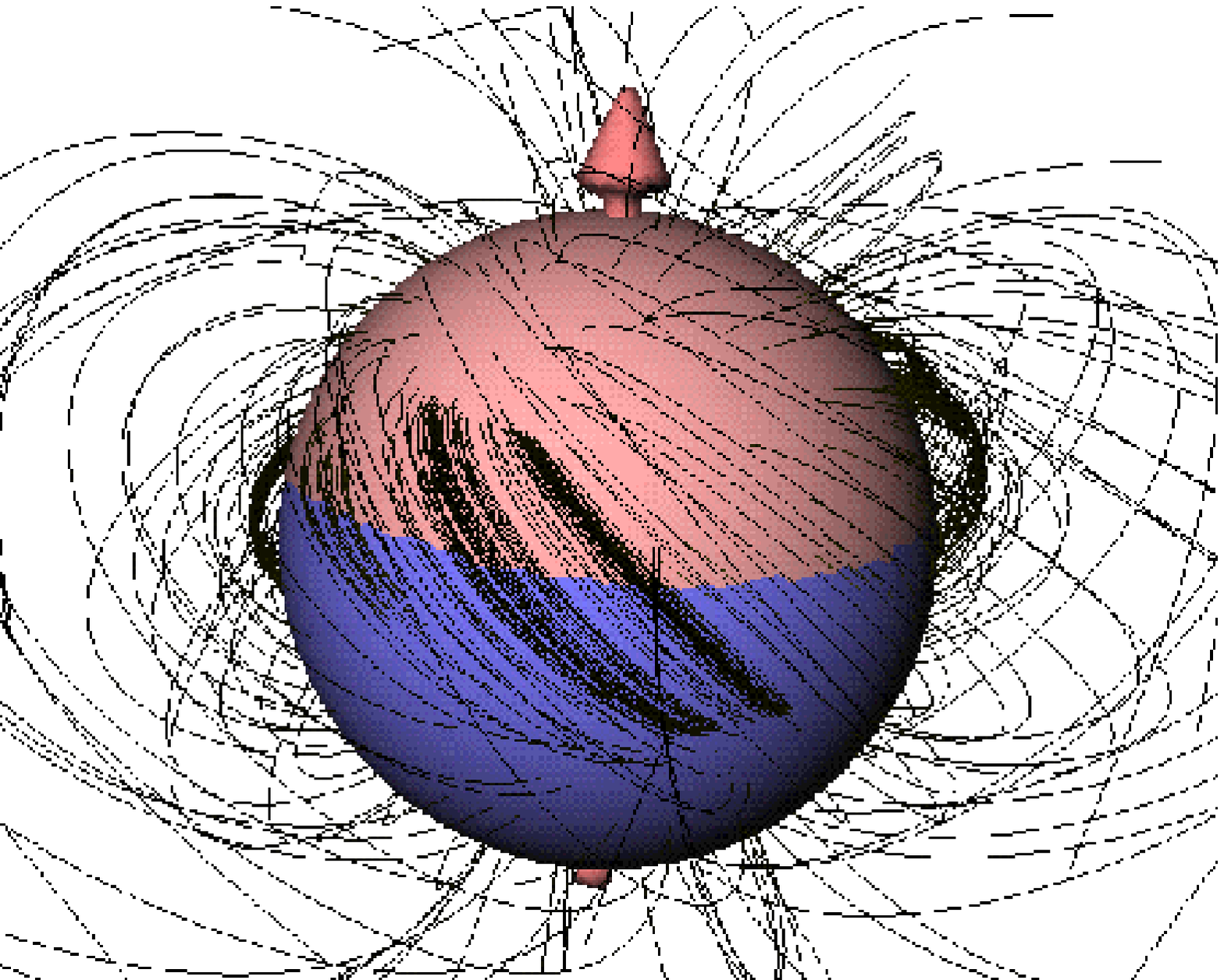}
\includegraphics[width=0.48\hsize]{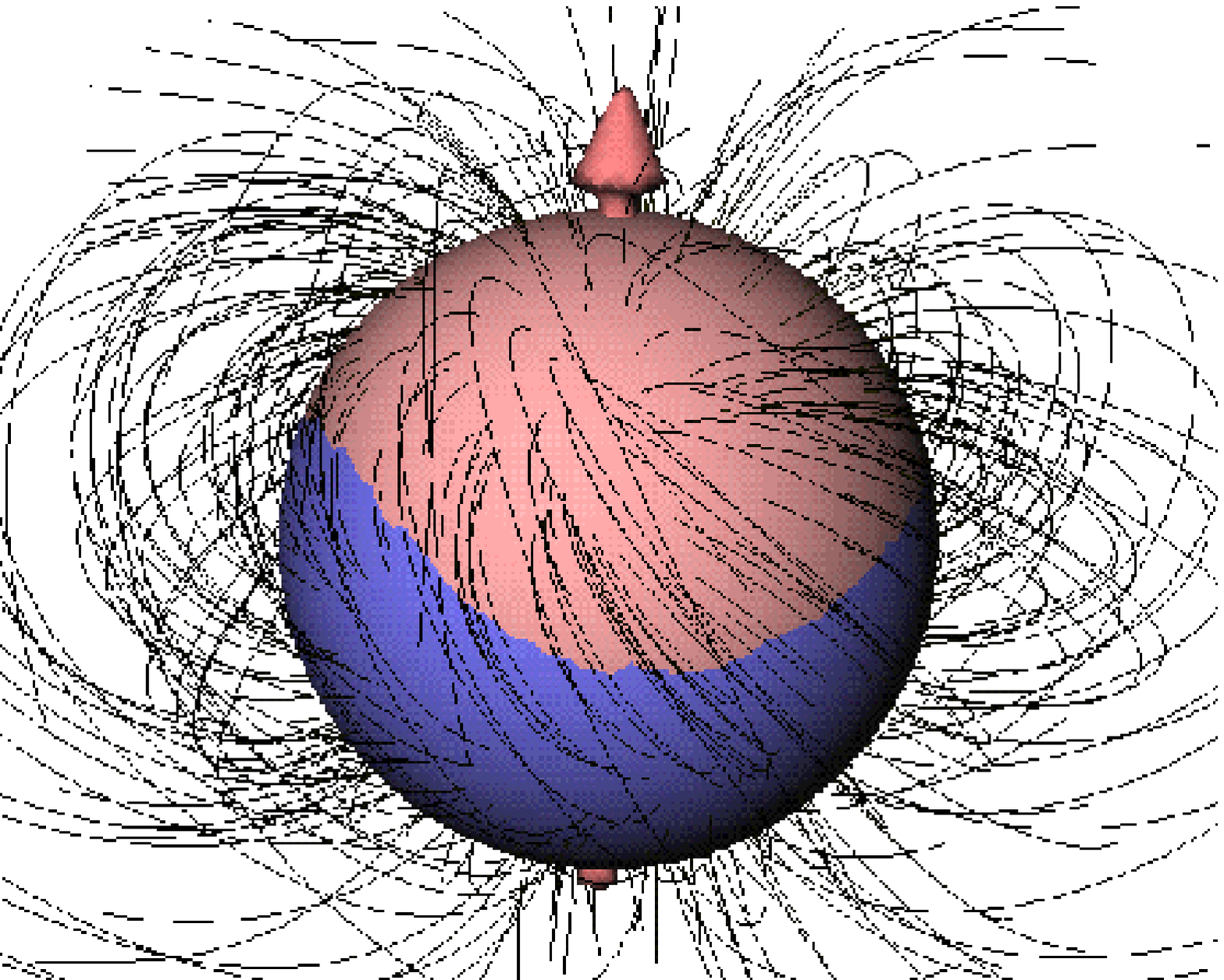}
\includegraphics[width=0.48\hsize]{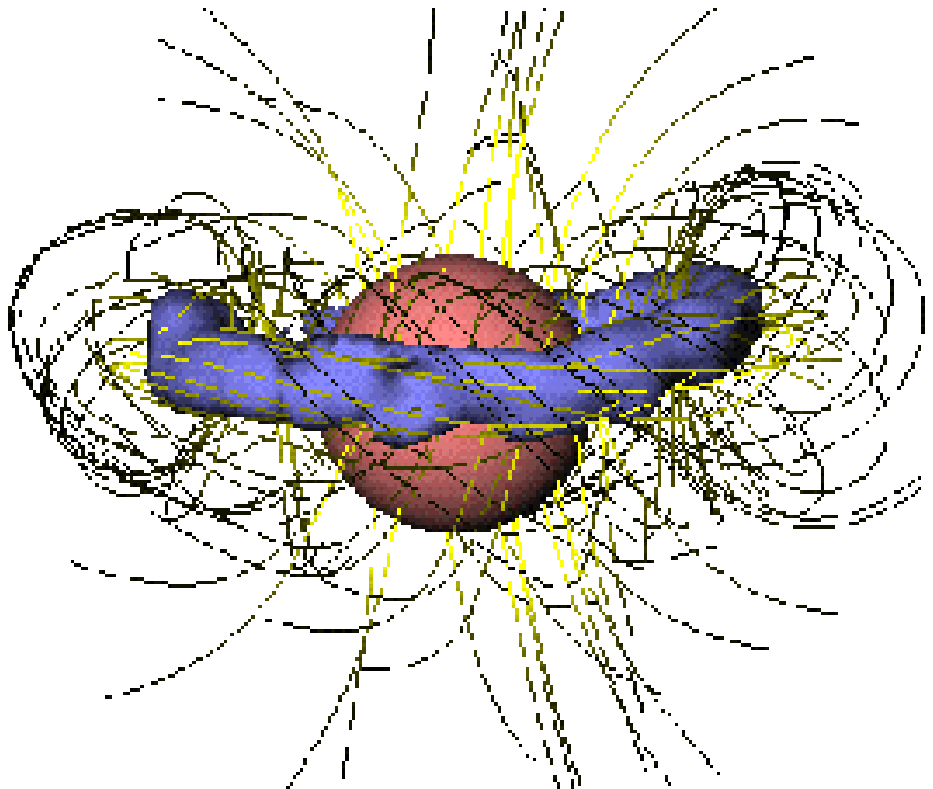}
\includegraphics[width=0.48\hsize]{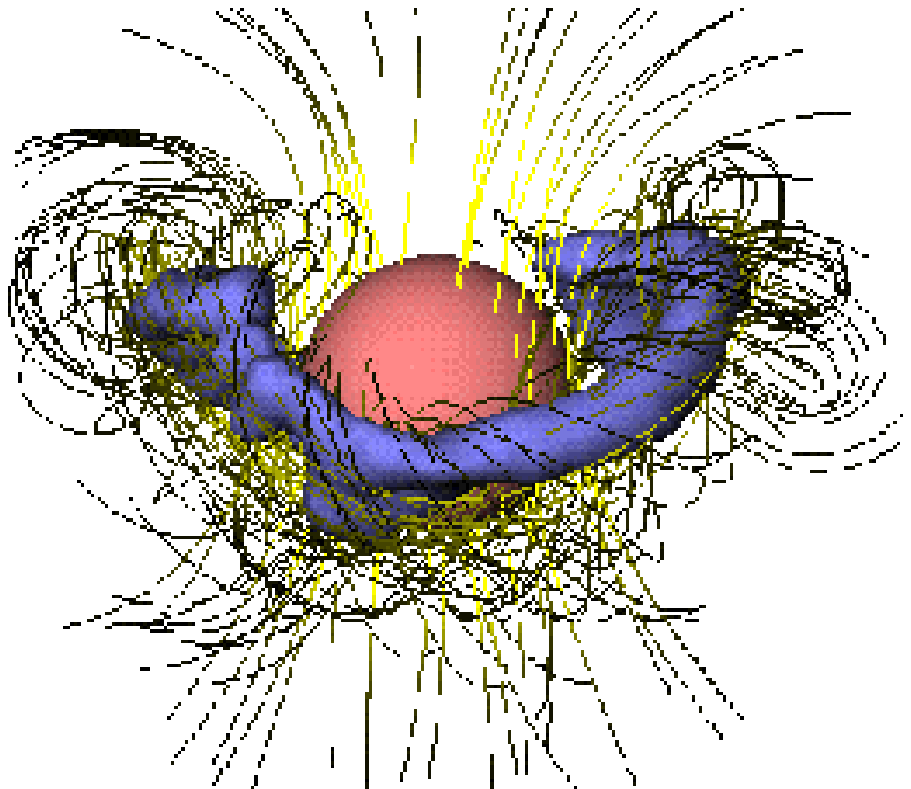}
\includegraphics[width=0.48\hsize]{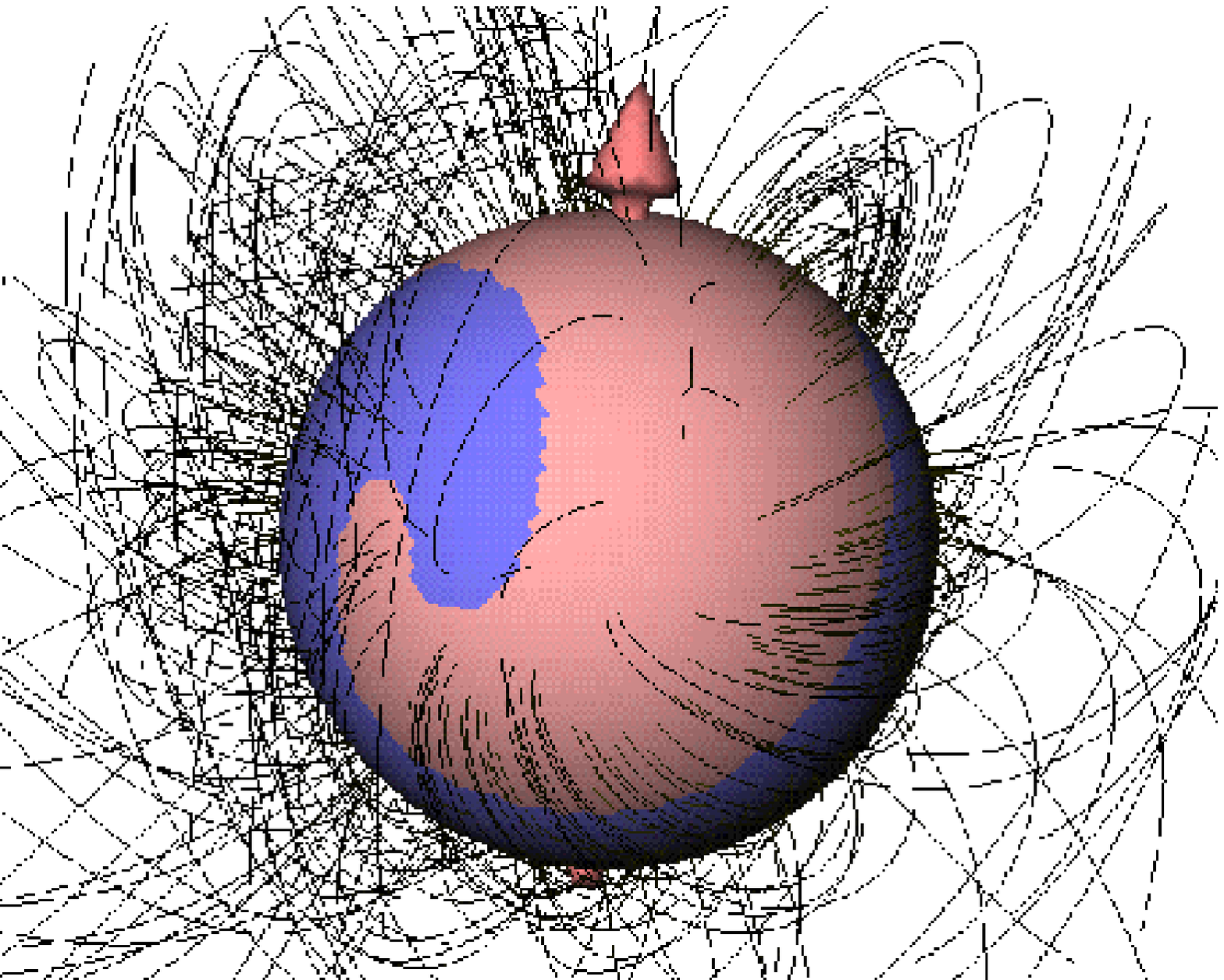}
\includegraphics[width=0.48\hsize]{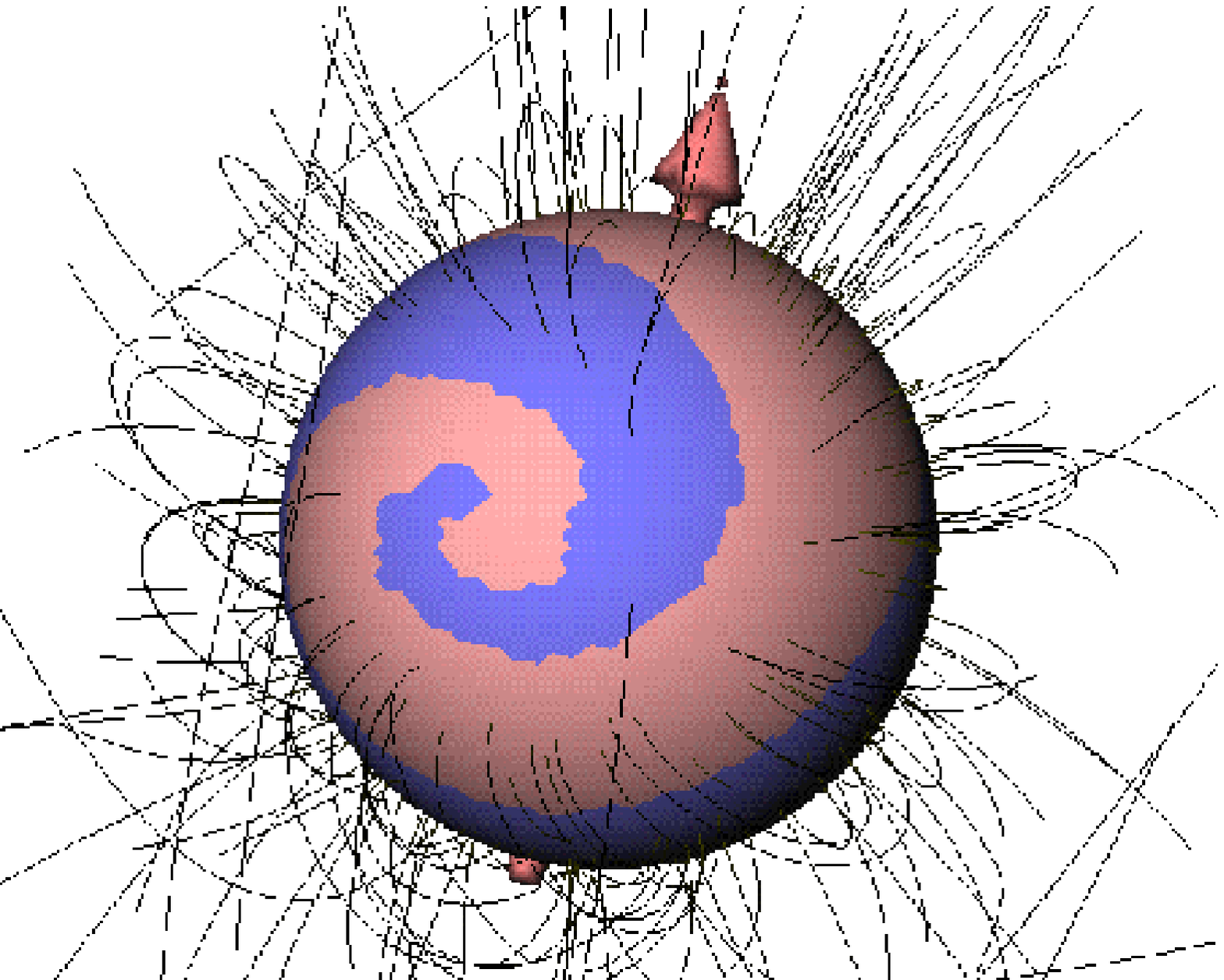}
\includegraphics[width=0.48\hsize]{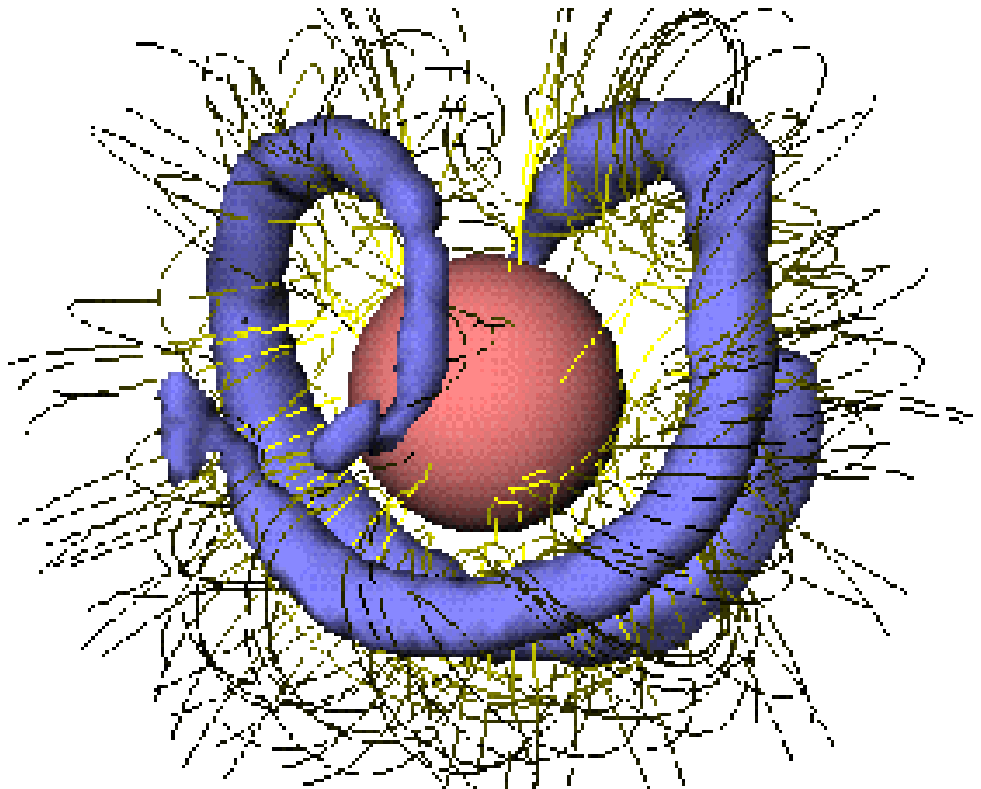}
\hfill
\includegraphics[width=0.48\hsize]{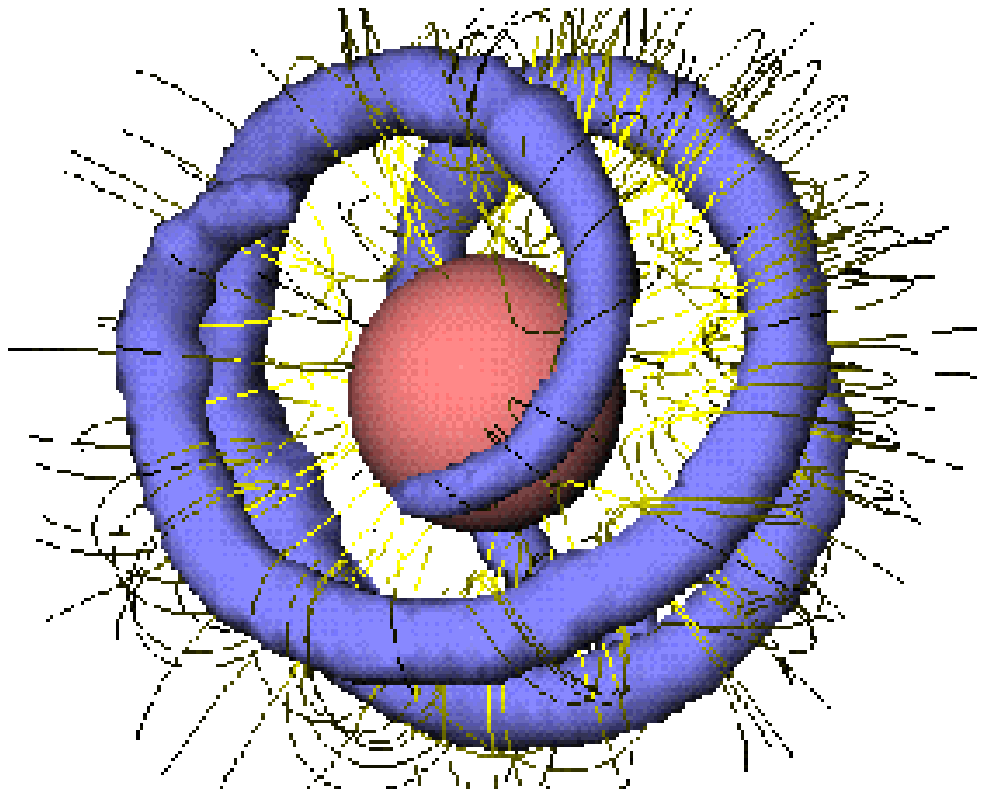}
\caption{Large plates: the magnetic field in the atmosphere of the star. The light
and dark shading on the surface represent areas of positive and negative
$B_r$, the radial component of the field.
The arrow denotes the magnetic axis $\mathbf{M}$ calculated in Eq. (\ref{eq:m1}).
The four snapshots are taken at times
$t=31.9$, $33.7$, $35.6$ and $38.5$ days;
top-left, top-right, bottom-left, bottom-right respectively. In the first, the field has settled from 
the initial state into a fairly regular circular torus. In the next three we can see the
instability grow (see Sect. \ref{sec:phase3}).
Small plates: at the same times, on the same scale, field lines in the
stellar interior. To make it easier to trace their
path, a surface of constant $G$ (see Sect. \ref{sec:vis} and
Eq. (\ref{eq:finaltorusplot})) has been added, as well as a sphere of
radius $0.3R_\ast$.}
\label{fig:foursnaps}
\end{figure*}

\begin{figure*}
\includegraphics[width=0.5\hsize]{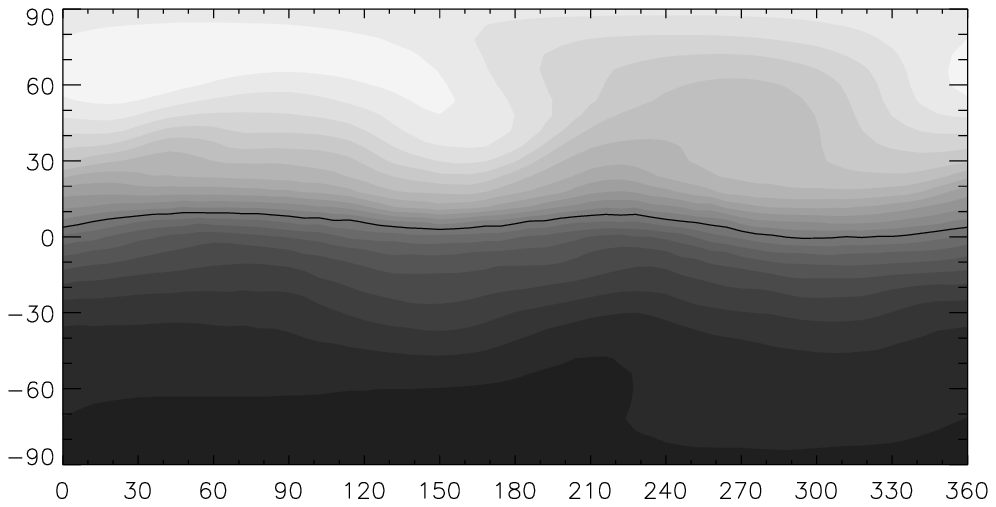}
\includegraphics[width=0.5\hsize]{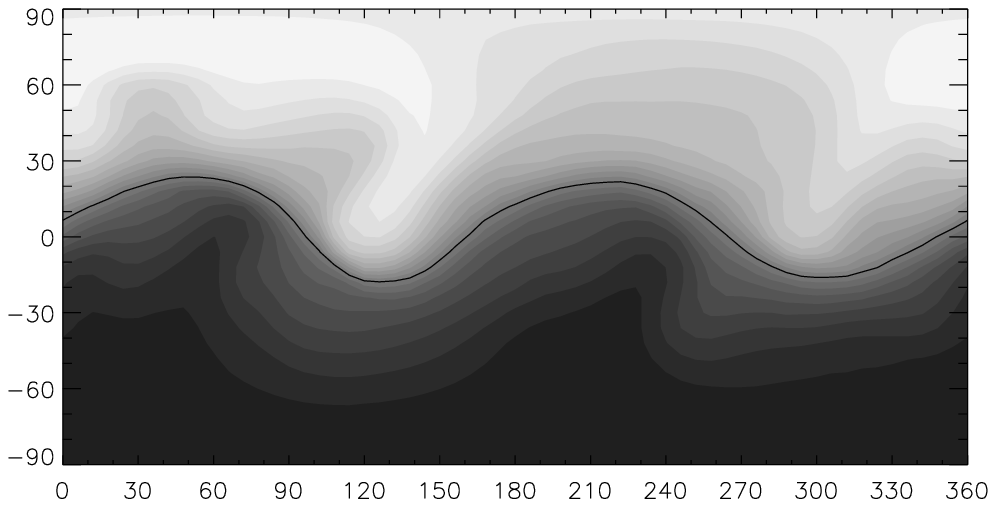}
\includegraphics[width=0.5\hsize]{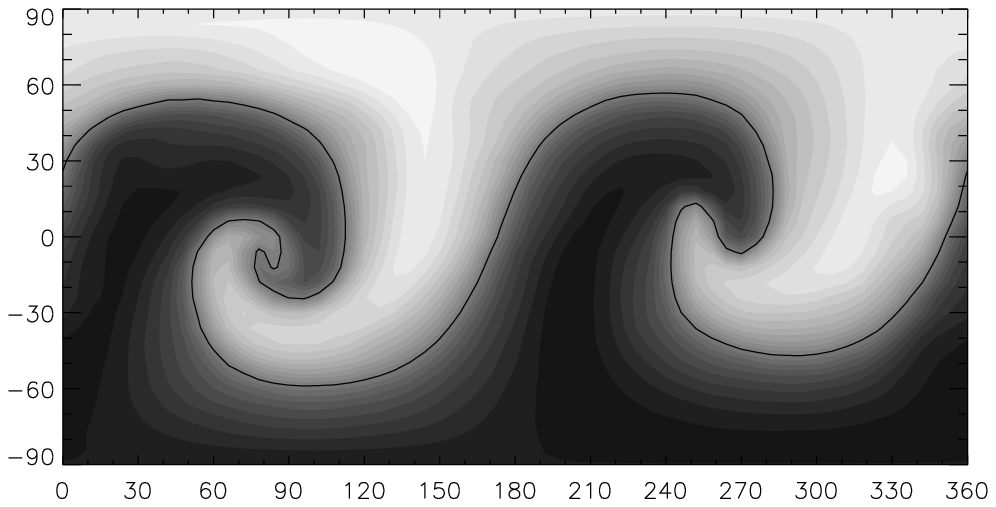}
\includegraphics[width=0.5\hsize]{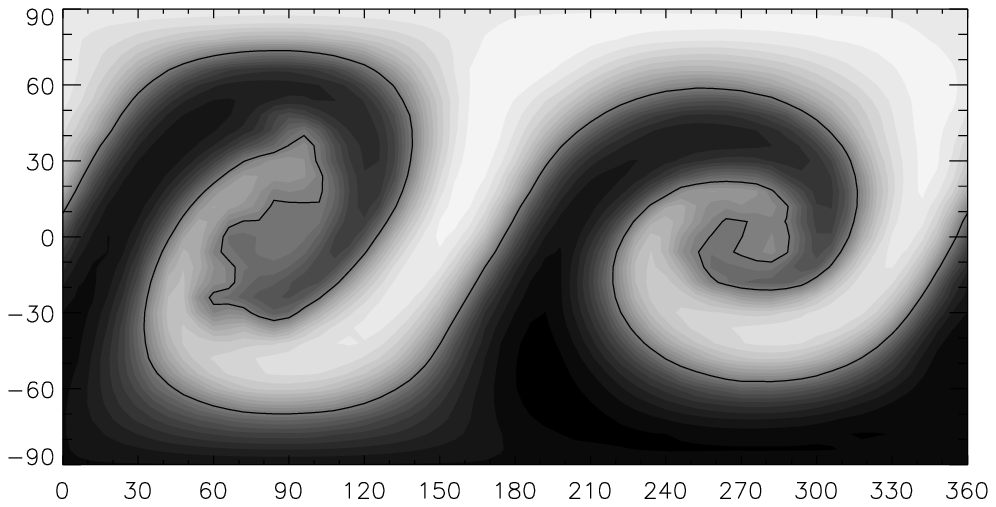}
\includegraphics[width=0.5\hsize]{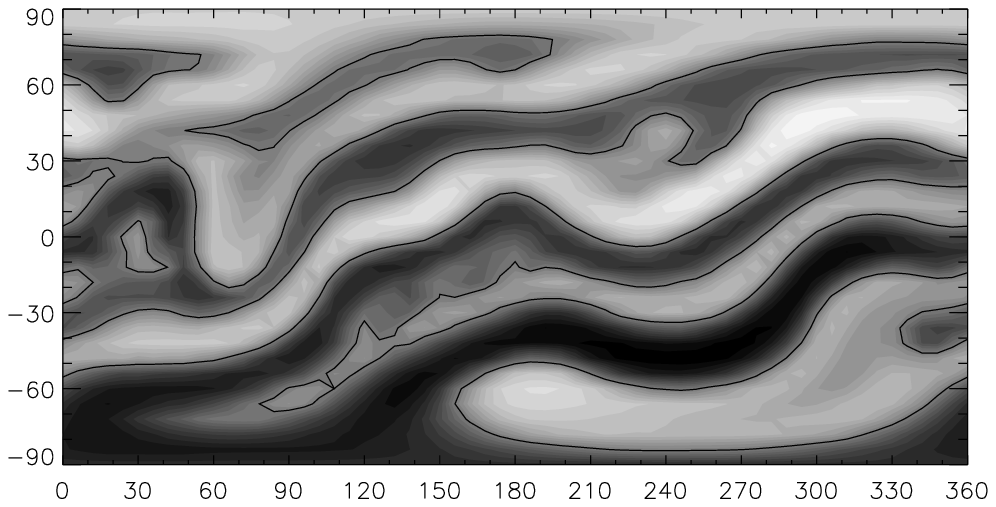}
\includegraphics[width=0.5\hsize]{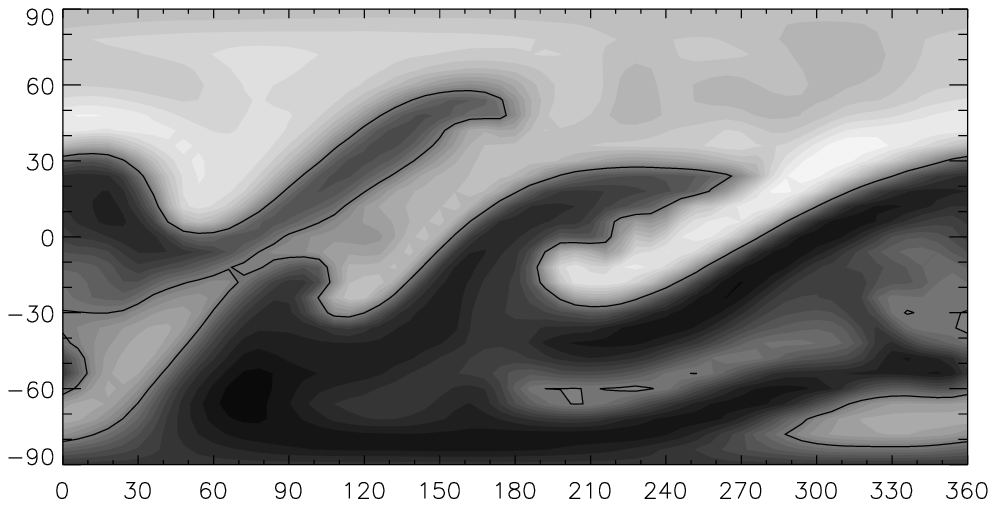}
\caption{Projections onto 2-D of the radial component $B_r$ on the
stellar surface, for the fiducial run at resolution $96^3$, at times
$t=31.9, 33.7, 35.6, 38.5, 43.2, 46.0$ days, using the
axis $\mathbf{M}$.}
\label{fig:2Dprojectphase3}
\end{figure*}

Fig.~\ref{fig:foursnaps} shows the evolution of this final instability, from
the moment when it begins to a time when the length scale has fallen
significantly. Fig.~\ref{fig:2Dprojectphase3} is a projection
onto two dimensions of the field's radial component $B_r$ on the
stellar surface. The axis $\mathbf{M}$ at the time of the third
picture ($t=31.9$ days) is used for the projection, although this axis moves 
by less than five degrees between then and the time of the last picture in the
sequence. The third, fourth, fifth and sixth frames in Fig. 
\ref{fig:2Dprojectphase3} correspond to the four frames in 
Fig. \ref{fig:foursnaps}.





To see the length scale falling, we can use the quantity $W$, which
was plotted in Fig.~\ref{fig:wiggliness}. At around $t=35$ days, $W$
suddenly rises, and since the length scale is given by
$\mathcal{L}\sim 2R_\ast/W$, i.e. the length scale is inversely proportional to $W$,
this means the length scale is going down.

As we saw in Fig.~\ref{fig:sh-comps} (which shows how the energy of
the field on the surface is divided up between the different
components in spherical harmonics), the field is almost entirely
dipolar just before the field becomes unstable. Then, the dipolar
component decays very rapidly and higher orders take over.

\begin{figure}
\includegraphics[width=1.0\hsize,angle=0]{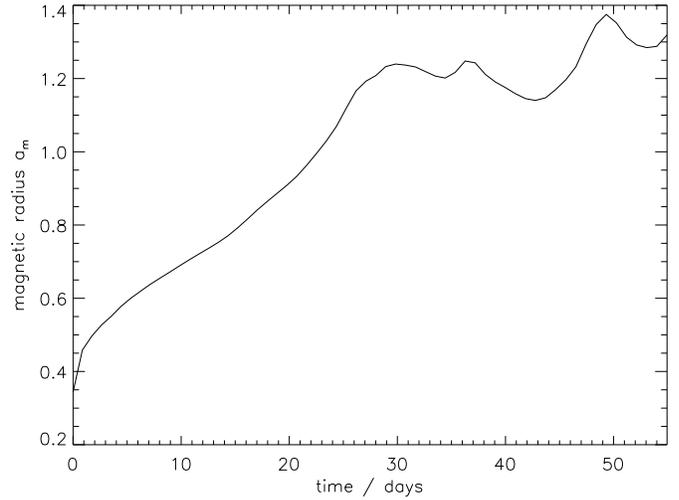}
\caption{The magnetic radius $a_\mathrm{m}$ against time. It can be
seen that during the diffusive phase, the field moves slowly outwards,
and that the field enters the unstable phase only once this outwards
diffusion has ended. This transition to instability, as can be seen for instance in Fig. \ref{fig:sh-comps},
occurs at $t\approx 35$ days.}
\label{fig:fid-am}
\end{figure}

The beginning of this unstable phase marks the end of the gradual
outwards diffusion of the field. This can be seen by looking at the
value of the magnetic radius $a_\mathrm{m}$ (defined in
Eq. (\ref{eq:def-am})), as plotted in Fig. \ref{fig:fid-am}.

\section{Comparison with observations}
\label{sec:compobs}

The main result of this study is the existence of a dynamically stable field which
can survive for a long time, and assuming the validity of the diffusivity extrapolation described in Sect.~\ref{sec:phase2} and Fig.~\ref{fig:phase2-etatsr}, at least as long as an A-star main
sequence lifetime. At the surface of the star, this field is found to be mainly 
dipolar, with smaller contributions from quadrupole and higher components. 
This is largely in agreement with the observations described in Sect.
\ref{sec:hist}; it is observed that the field on the surface of Ap
stars is ordered on a large scale and mainly dipole in form. 

We should like to make this comparison between the result of this study and
the observations in a slightly more quantitative manner. To this end we
can calculate from the results found here some quantities which can be 
directly observed. This bypasses the processes involved in reconstructing 
the surface field from observations.

The most common method for looking at the magnetic field of
a star is the analysis of the Stokes $I$ and $V$ profiles and of
frequency-integrated Stokes $Q$ and $U$ profiles. This can yield various
quantities depending on factors such as the rotation velocity
of the star (which broadens the spectral lines and therefore makes the
analysis more difficult). The most easily obtained of these quantities
are the {\it longitudinal field} $B_l$ and the {\it field modulus} $B_s$, 
which are obtained from weighted averages over the visible hemisphere
of the line-of-sight component $\langle B_z \rangle$ and of the modulus 
$\langle B \rangle$ of the field respectively. They are weighted with the
function $Q(\Theta)$, where $\Theta$ is the angle between the
normal to the stellar surface and line of sight, which is given by (\cite{LanandMat:2000})
\begin{equation}
Q(\Theta)=[1-\epsilon_c(1-\cos\Theta)][1-\epsilon_l(1-\cos\Theta)],
\end{equation}
where $\epsilon_c$ and $\epsilon_l$ are limb darkening and line
weighting coefficients, given the values $0.4$ and $0.5$.

As the star rotates, these quantities $B_l$ and $B_s$ change; we can calculate them as
a function of rotational phase. We just need to choose a rotational
axis, and an angle $i$ between this rotational axis and the line of
sight.
The matter of whether the magnetic and rotation axes are likely to
be close together or far apart is a little uncertain. It is possible
that rotation has a direct effect on the direction of the magnetic
axis, or that the magnetic field makes the rotation axis precess and migrate
(\cite{MesandTak:1972} and references therein) --
whether this tends to increase or decrease the angle between the two
depends on the shape of the field. For the slow rotators though, it
seems that the two are close together (\cite{LanandMat:2000}), so we shall pick an angle $30^\circ$.
For $i$, we choose the median value (the distribution
is of course random) of $60^\circ$.

\begin{figure}
\includegraphics[width=0.48\hsize,angle=0]{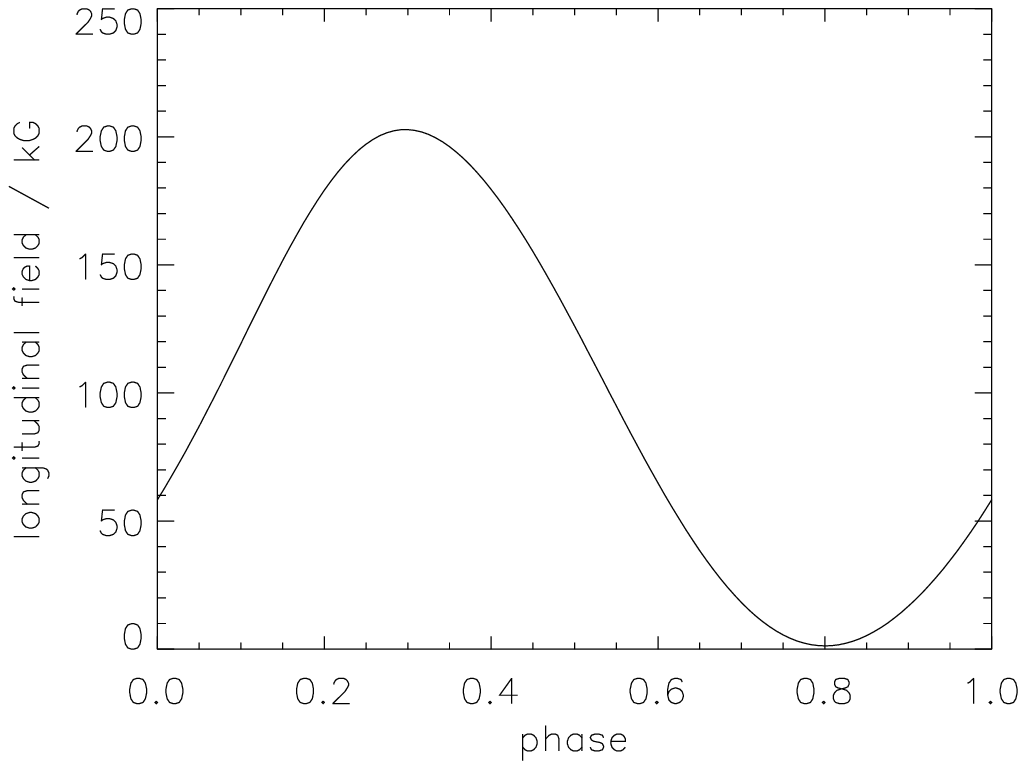}\hfill
\includegraphics[width=0.48\hsize,angle=0]{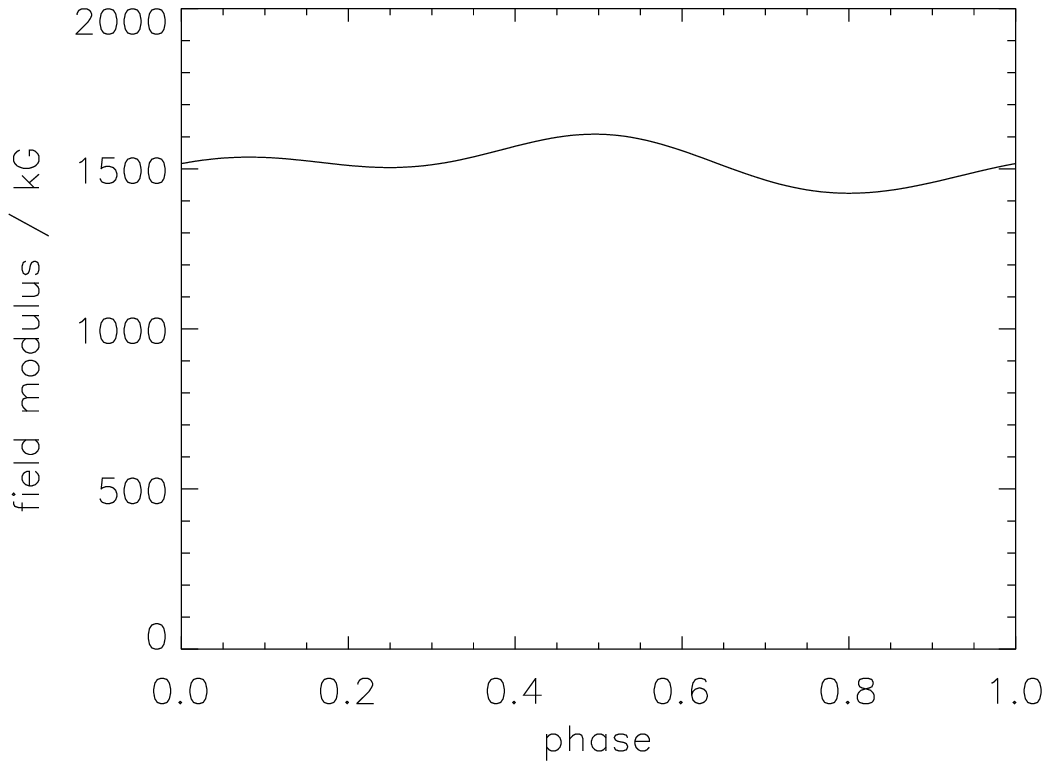}\\
\includegraphics[width=0.48\hsize,angle=0]{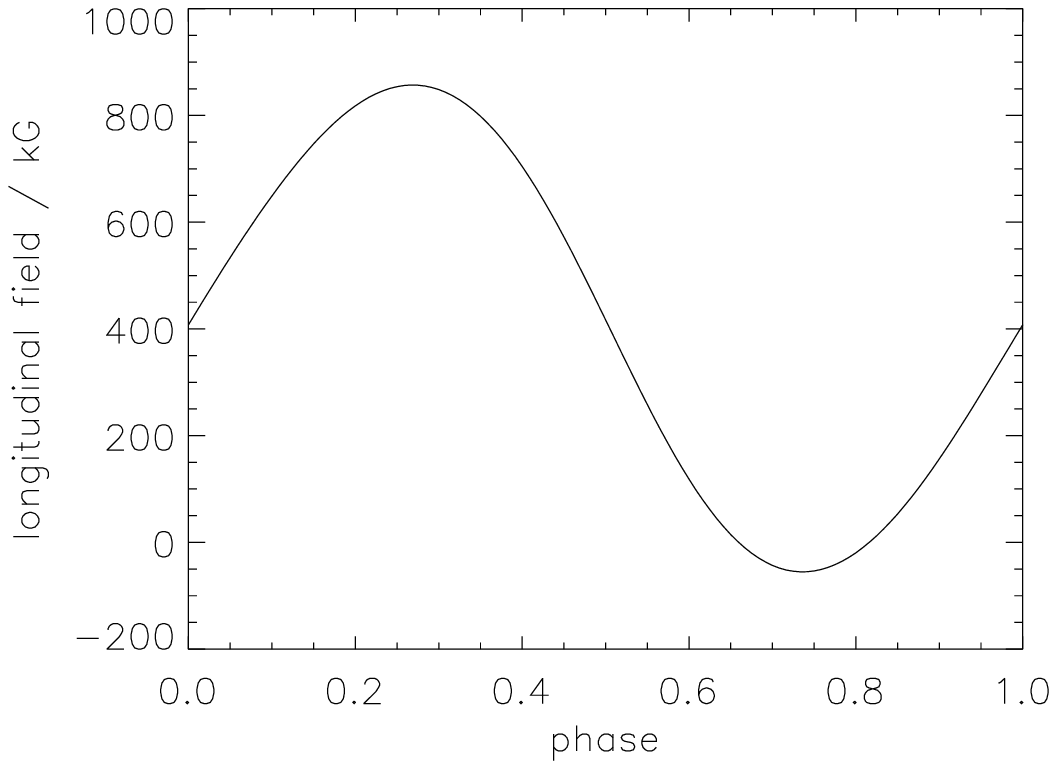}\hfill
\includegraphics[width=0.48\hsize,angle=0]{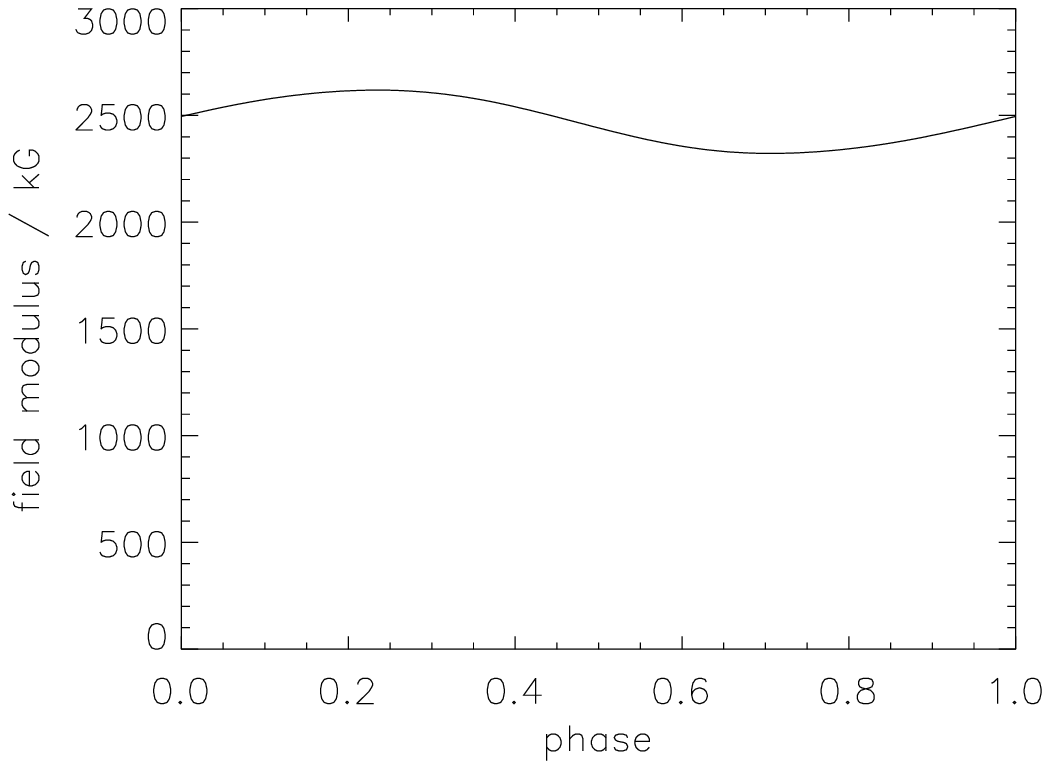}
\caption{\label{fig:observations} Longitudinal field and field modulus
at two stages during the field's evolution, as seen over one
rotation. Upper left frame: longitudinal field at $t=22.6$ days; upper
right: field modulus at $t=22.6$ days; lower left: longitudinal field
at $t=31.9$ days, lower right: field modulus at $t=31.9$ days}
\end{figure}

We calculate the longitudinal field and field modulus from the 
fiducial run described in Sect. \ref{sec:results}. We have done this at two
points in time while the stable torus field is present, $t=22.6$ and
$31.9$ days (corresponding to the second and third frames of 
Fig. \ref{fig:2Dprojectphase3}), the surface field being significantly stronger 
at the later time. These quantities are plotted, as a function of rotational
phase, in 
Fig.~\ref{fig:observations}.

At the later of these two times, both the longitudinal field and the
field modulus are variable in a sinusoidal fashion, which is what is
observed in most Ap stars. At the earlier time $t=22.6$ days, the
variation is not purely sinusoidal. This behaviour has been observed
for instance in the slowly rotating Ap star HD 187474
(\cite{Khalacketal:2003}).

\section{Discussion and conclusions}
\label{sec:disc}

\begin{figure*}
\includegraphics[width=1.0\hsize,angle=0]{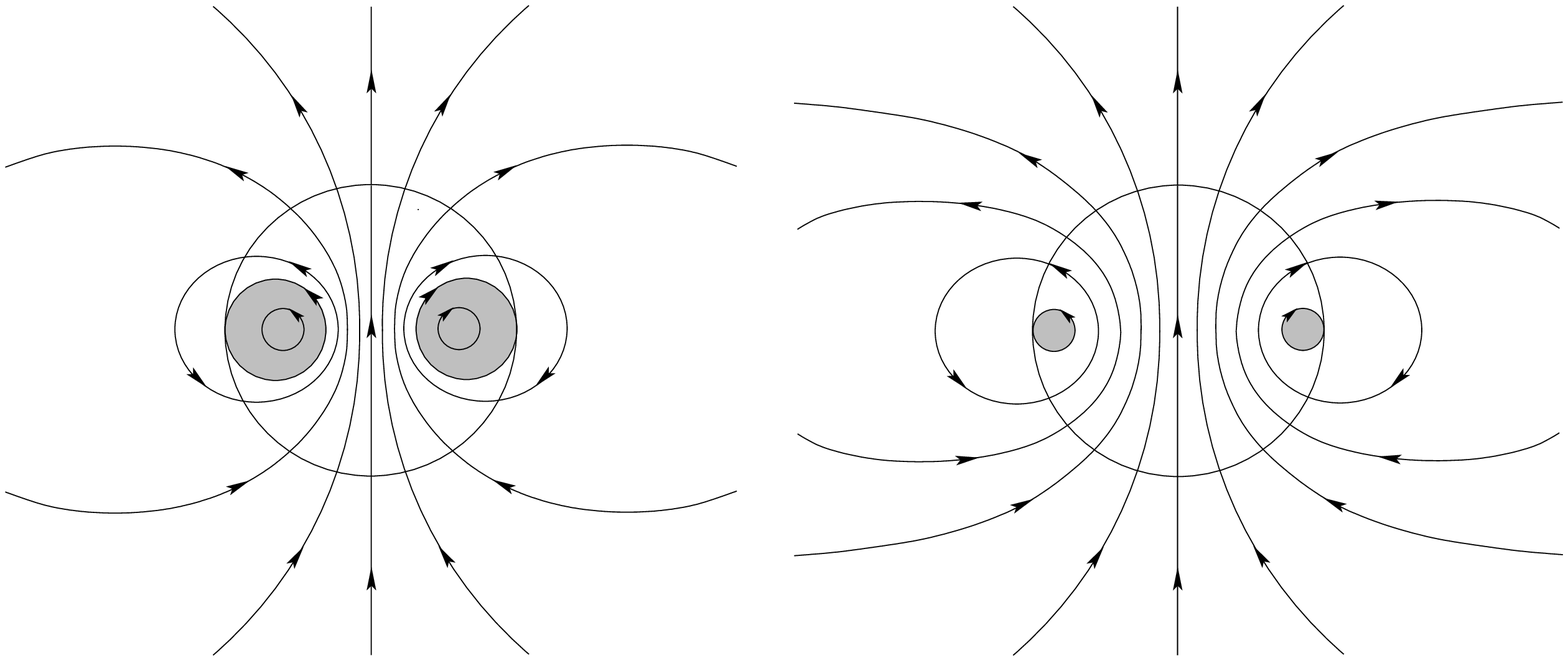}
\caption{Left: The toroidal field lines (represented by the shaded
area) thread through those poloidal field lines which are closed within
the star. Right: at a later time the field has diffused outwards and
the toroidal component has been reduced compared to the poloidal.}
\label{fig:tortopol}
\end{figure*}

We have modelled an Ap star and evolved its magnetic field in time using
numerical MHD, starting with a random field configuration in the
interior of the star. Any random field configuration is in general
unstable and will decay on a timescale comparable to the time taken by
an Alfv\'{e}n wave to cross the star; this is indeed exactly what
happens in these simulations. The field evolves into a stable `twisted
torus' configuration, which is then stable on an Alfv\'en time-scale
(dynamic stability). 

The configurations found were always of the same kind: a
nearly axisymmetric torus inside the star, with toroidal (azimuthal) and
poloidal (meridional) components of comparable strength. Depending on 
the particular random field present at the beginning, the torus which 
emerges is either right or left handed, and is in general a little displaced 
from the centre of the star. This torus
forms the stable core of the configuration. Wrapped around it are poloidal
field lines that extend through the atmosphere. These field lines cause the
surface field to form an approximate dipole, with smaller contributions
from higher multipoles. 

The first conclusion to be drawn from this study is therefore the probable 
existence of stable field configurations in stably stratified stars. 
Our results provide the first plausible field configurations that explain both 
the stability and the surface appearance of A-star fields. We consider the 
results to be strong evidence in favour of the `fossil field' model for Ap stars.

The second main result concerns the secular evolution of the stable
field configurations. 
By Ohmic dissipation the field diffuses slowly outwards, while maintaining 
its overall shape. In the process, the strength of the field on the surface 
{\it increases} and the topology of the field in the interior undergoes a 
gradual change, from mainly toroidal to mainly poloidal. To understand 
why this happens, one first needs to understand that the toroidal field
can only thread through those poloidal field loops which are closed
inside the star. If toroidal field were present in regions where the
poloidal field lines go all the way through the star and close
outside, the field lines would be in effect entering the star at the
north pole, twisting around inside the star, and exiting at the south
pole. 
Due to the rapid relaxation of the atmospheric field, this twist is removed
almost instantaneously, such that the toroidal field component is always
small outside the star. As a result, the field line does not support a torque
any more, and the interior part of the field line unwinds on an Alfv\'en
time-scale (by azimuthal displacements) until only its poloidal component
remains. This is sketched in Fig. \ref{fig:tortopol}. As seen in a projection 
on a meridional plane, the toroidal field component is restricted to those
field lines that are closed within the star. As these closed field lines diffuse
to the surface, their toroidal component is released, increasing the ratio
of poloidal to toroidal field energy. 

At some point, when the torus of closed field lines is close to the surface 
of the star, the (relative) increase of the poloidal field causes the torus to
loose its circular shape. 
Its core, now located just below the surface, twists out of the plane.
At first, this twist is like the seams on a tennis ball and thereafter 
becomes increasingly complicated. This continues
until the diffusive timescale $\mathcal{L}^2/\eta$ falls to the
Alfv\'{e}n time-scale. The field then decays on an extremely short
timescale compared to the lifetime of the star.

The diffusive evolution of the field thus agrees with the observational
result in Sect. \ref{sec:hist} of Hubrig et al. (2000a), that Ap 
stars are typically more than $30\%$ of the way through their main-sequence 
life. The time-scale 
for this increase is found to be around $2\times 10^9$ years, somewhat
longer than $30\%$ of the main-sequence lifetime of an A star ($30\%$
of $10^9$ years), but it should be stressed that any accurate
determination of this diffusive time-scale would have to use a more
accurate modelling of the stellar structure and the magnetic diffusivity 
than that employed here. 

For our results to hold as an explanation of A-star magnetic fields, the stars 
must, at the time of their formation, have contained a strong field. This initial 
field can be of arbitrary configuration, except that it must have a finite magnetic 
helicity and must be confined mainly to the core. Why only some A stars show 
a strong field is another question. There is of course the obvious possibility 
that the $\sim 90\%$ of A stars not observed to be magnetic simply contained, for
whatever reason, no strong field at birth. There are however two other
possible reasons, albeit also with no obvious explanation for the birth state
required. Firstly, it is possible that most A stars are born with a
strong field which is not sufficiently concentrated into the core, so
that it quickly or immediately becomes unstable and decays, analogous
to the run with $r_\mathrm{m}=0.57R_\ast$ described in
Sect. \ref{sec:rm} and Figs. \ref{fig:rmeffect-pc} and \ref{fig:rmeffect-nd}. 
Secondly, the field in most A stars could be more concentrated towards the
centre than in Ap stars, so that it does not have time to manifest itself at the surface during 
the main sequence.

It may seem unnatural that a newly born star should have its magnetic field
concentrated into the core. However, this may be a logical consequence of
flux conservation during formation (see, for instance,
\cite{MesandSpi:1956}, \cite{Mestel:1966}). Assuming that the field in the
progenitor cloud is of uniform strength, and that the topology of the
field does not change, the field strength in the newly formed star will be
proportional to $\rho^{2/3}$. In a polytrope of index $3$, as we have
used for this model, the ratio of thermal to magnetic energy
densities $\beta=8\pi e\rho/B^2$ will be constant, independent of
radius. This is roughly the situation we have in the fiducial run in
this study with $r_\mathrm{m}=0.25R_\ast$ (see Fig. \ref{fig:profile-enden}), 
and that is indeed sufficiently concentrated to produce a stable torus field. This is assuming that the field the main-sequence star is born with comes directly from the cloud it formed from. There is of course another possibility: that of a dynamo in the pre-MS star which uses the cloud field merely as a seed. The magnetic field energy in this case is however also likely to be concentrated where the density is highest, i.e. in the centre.

There are examples of binary systems containing both a magnetic A star 
and a non-magnetic A star. This rules out chemical composition ({\it overall} chemical compostion, not the composition we see on the surface, which is the result itself of the magnetic field) as the 
reason for the difference. It perhaps also rules out the field strength in the cloud
from which the star condenses. It does not rule out the initial
angular momentum distribution, and the effect this may have on any kind of
dynamo driven by differential rotation; indeed it is possible that
rotation has an effect on whether the field emerges to be seen at the surface
(\cite{MesandMos:1977}). Nor does it rule out differences in 
the precise shape of the field in the 
accretion discs that feed the growing protostar. Some fields of
random shape will find their way to the stable configuration faster
than others, losing less magnetic flux in the process. This may also
have an effect on the size of the torus produced, which may then, as
described in the previous paragraph, determine whether any field is
observed on the surface.

\begin{appendix}

\end{appendix}


\begin{flushleft}

\end{flushleft}

\end{document}